\documentclass[12pt]{article}
\usepackage[hidelinks]{hyperref}

\usepackage{graphicx}           
\usepackage{amsfonts}            
\usepackage{amssymb}             
\usepackage{amsthm}              
\usepackage{amsmath}            
\usepackage{rotating}            
\usepackage{xspace}
\usepackage[margin=1in]{geometry}
\usepackage{multirow}
\usepackage{hyphenat}		
\usepackage{paralist}       

\usepackage{todonotes}      
\usepackage{listings} 
\usepackage[T1]{fontenc} 

\newcommand{\cmar}[1]{} 
\renewcommand{\cmar}[1]{\todo[linecolor=red,color=green!40]{#1 \\ \texttt{\scriptsize CSM}}} 

\usepackage{color}
\newcommand{\cinl}[1]{} 

\usepackage{setspace}                
\doublespace                        

\usepackage{natbib}
\bibpunct{(}{)}{;}{a}{,}{;}

\newcommand{\citepos}[1]{}
\renewcommand{\citepos}[1]{\citeauthor{#1}'s (\citeyear{#1})}

\begin{document}
\title{A Relational Event Approach to Modeling Behavioral Dynamics}

\author{Carter T. Butts \thanks{Departments of Sociology, Statistics, and EECS and Institute for Mathematical Behavioral Sciences, University of California---Irvine; \texttt{buttsc@uci.edu}} \and Christopher Steven Marcum \thanks{National Institutes of Health; \texttt{chris.marcum@nih.gov}}}
\date{\today{}}
\maketitle


\begin{quote}
DISCLAIMER: This self-archived version of the paper is provided by the authors after publication. The authors never received proofs to correct from Springer prior to printing.  Minor corrections and proper typesetting have been committed to this version of the manuscript. 

The proper citation for this paper is:

\begin{verbatim}
Butts, Carter T and Christopher Steven Marcum. 2017. ``A Relational 
   Event Approach to Modeling Behavioral Dynamics." In Andrew Pilny and 
   Marshall Scott Poole, eds., Group Processes: Data-Driven 
   Computational Approaches, pages 51-92. Springer International 
   Publishing: Cham, Switzerland.
\end{verbatim}

This paper is cross-deposited in both arXiv and SocArXiv. The referenced example dataset and R script can be downloaded from both repositories.
\end{quote}

\bigskip

\section{Representing Interaction: from Social Networks to Relational Events}

The social network paradigm is founded on the basic representation of social structure in terms of a set of social entities (e.g., people, organizations, or cultural domain elements) that are, at any given moment in time, connected by a set of relationships (e.g., friendship, collaboration, or association) \citep{wasserman.faust:bk:1994}.  The success of this paradigm owes much to its flexibility: with substantively appropriate definitions of entities (\emph{vertices} or \emph{nodes} in network parlance) and relationships (\emph{ties} or \emph{edges}), networks can serve as faithful representations of phenomena ranging from communication and sexual relationships to neuronal connections and the structure of proteins \citep{butts:s:2009}.  Nor must networks be static: the time evolution of social relationships has been of interest since the field's earliest days \citep[see, e.g.][]{heider:jp:1946,rapoport:bmb:1949a,sampson:bk:1969}, and considerable progress has been made on models for network dynamics \citep[e.g.][]{snijders:sm:2001,koskinen.snijders:jspi:2007,almquist.butts:sm:2014,krivitsky.handcock:jrssB:2014}.  Such models treat relationships (and, in some cases, the set of social entities itself) as evolving in discrete or continuous time, driven by mechanisms whose presence and strength can be estimated from intertemporal network data.

A key assumption that underlies the network representation in both its static and dynamic guises is that relationships are \emph{temporally extensive}---that is, it is both meaningful and useful to regard individual ties as being present for some duration that is at least comparable to (and possibly much longer than) the time scale of the social process being studied.  Where tie durations are much longer than the process of interest, we may treat the network as effectively ``fixed;'' thus is it meaningful for \citet{granovetter:ajs:1973} or \citet{burt:bk:1992} to speak of personal ties as providing access to information or employment opportunities, for \citet{friedkin:bk:1998} to model opinion dynamics in experimental groups, or for \citet{centola.macy:ajs:2007} to examine the features that allow complex contagions to diffuse in a community, without explicitly treating network dynamics.  When social processes (including tie formation and dissolution themselves) occur on a timescale comparable to tie durations, it becomes vital to account for network dynamics.  For instance, the diffusion of HIV through sexual contact networks is heavily influenced by partnership dynamics (particularly the formation concurrent rather than serial relationships) \citep{morris.et.al:ch:2007}, and health behaviors such as smoking and drinking among adolescents are driven by an endogenous interaction between social selection and social influence \citep[see, e.g.][]{lakon2015simulating,wang.et.al:pab:2016}.  While there are many practical and theoretical differences between the behavior of networks in the dynamic regime versus the ``static'' limit, both regimes share the common feature of \emph{simultaneity}: relationships overlap in time, allowing for apparent reciprocal interaction between them. 

Such simultaneous co-presence of edges forms the basis of all network structure (as expressed in concepts ranging from reciprocity and transitivity to centrality and structural equivalence), and is the foundation of social network theory.  Such simultaneity, however, is a hidden consequence of the assumption of temporal extensiveness; in the limit, as tie durations become much shorter than the timescale of relationship formation, we approach a regime in which ``ties'' become fleeting interactions with little or no effective temporal overlap.  In this regime the usual notion of network structure breaks down, while alternative concepts of \emph{sequence} and \emph{timing} become paramount.

This regime of social interaction is the domain of \emph{relational events} \citep{butts2008rems}.  Relational events, analogous to edges in a conventional social network setting, are discrete instances of interaction among a set of social entities.  Unlike temporally extensive ties, relational events are approximated as instantaneous; they are hence well-ordered in time, and do not form the complex cross-sectional structures characteristic of social networks.  This lack of cross-sectional structure belies their richness and flexibility as a representation for interaction dynamics, which is equal to that of networks within the longer-duration regime.  (In fact, the two regimes can be brought together by treating relationships as spells with instantaneous start and end events.  Our main focus here is on the instantaneous action case, however.)  The relational event paradigm is particularly useful for studying the social action that lies beneath (and evolves within) ongoing social relationships.  In this settings, relational events are used to represent particular instances of social behavior (e.g., communication, resource transfer, or hostility) exchanged between individuals.  To understand how such behaviors unfold over time requires a theoretical framework and analytic foundation that incorporates the distinctive properties of such micro-behaviors.  Within the relational event paradigm, actions (whether individual or collective) are treated as arising as discrete events in continuous time, whose hazards are potentially complex functions of the properties of the actors, the social context, and the history of prior interaction itself \citep{butts2008rems}.  In this way, the relational event paradigm can be viewed as a fusion of ideas from social networks and allied theoretical traditions such as agent-based modeling with the inferential foundation of survival and event history analysis \citep{mayer.tuma:bk:1990,blossfeld.rohwer:bk:1995}.  The result is a powerful framework for studying complex social mechanisms that can account for the heterogeneity and context dependence of real-world behavior without sacrificing inferential tractability. 

\subsection{Prefatory Notes}

At its most elementary level, as \citet{marcum&butts2015csse} point out, the relational event framework helps researchers answer the question of ``what drives what happens next'' in a complex sequence of interdependent events.  In this chapter, we briefly review the relational event framework and basic model families, discuss issues related to data selection and preparation, and demonstrate relational event model analysis using the freely available software package \texttt{relevent} for \texttt{R} \citep{butts2010rrem}.  Here, we provide some additional context before turning to the data and tutorial. 

Following \cite{butts2008rems}, a relational event is defined as an action emitted by one entity and directed toward another in its environment (where the entities in question may be sets of more primitive entities (e.g., groups of individuals), and self-interactions may be allowed).  From this definition, a relational event is thus comprised of a sender of action, a receiver of that action, and a type of action, with the action occurring at a given point in time.  In the context of a social system, we consider relational events as ``atomic units'' of social interaction.  A series of such events, ordered in time, comprise an event history that records the sequence of social actions taken by a set of senders and directed to a set of receivers over some window of observation.  The set of senders and the set of receivers may consist of human actors, animals, objects or a combination of different types of actors. The set of action types, likewise, may consist of a variety behaviors including communication, movements, or exchanges.  

The relational event framework is in an increasingly popular approach to the analysis of relational dynamics and has been adopted by social network researchers in a wide variety of fields.  Typically, research questions addressed in this body of work focus on understanding the behavioral dynamics of a particular type of action (such as communication alone).  Recently, relational event models have been used to study phenomena as diverse as reciprocity in food-sharing among birds \citep{tranmer&2015urem}; social disruption in herds of cows \citep{patison2015toea}; cooperation in organizational networks \citep{leenders3&2015outu}; conversational norms in online political discussions \citep{liang2014opod}; and multiple event histories from classroom conversations \citep{dubois&2013hmre}.  

Prior to the relational event framework, behavioral dynamics occurring within the context of a social network were generally modeled using frameworks developed for dynamic network data; since, as noted above, dynamic networks are founded on the notion of simultaneous, temporally extensive edges, use of dynamic network models for relational event data requires aggregation of events within a time window. Such aggregation leads to loss of information, and the results of subsequent analyses may depend critically on choices such as the width of the aggregation window.  Model families such as the stochastic actor-oriented models \citep{snijders1996saom} or the temporal exponential random graph models \citep{robins&pattison2001rgmt,almquist.butts:sm:2014,krivitsky.handcock:jrssB:2014} are appropriate for studying systems of simultaneous relationships that evolve with time, but may yield misleading results when fit to aggregates of relational events.  While such use can be motivated in particular cases, we do not as a general matter recommend coercing event processes into dynamic network form for modeling purposes.  Rather, where possible, we recommend that relational event processes be treated on their own terms, as sequences of instantaneous events with relational structure.  In the following sections, we provide an introduction to this mode of analysis.

\section{Overview of the Relational Event Framework}

We begin our overview of the relational event framework by considering what a relational event process entails.  Although we provide some basic notation, we omit most technical details; interested readers are directed to \citet{butts2008rems}, \citet{dubois&2013hmre}, and \citet{marcum&butts2015csse} for foundations and further developments.  We start with a set of potential senders, $S$, a set of potential receivers, $R$, and a set of action types, $C$.  A ``sender'' or ``receiver'' in this context may refer to a single individual or a set thereof; in some cases, it may be useful to designate a single bulk sender or receiver to represent the broader environment (if, e.g., some actions may be untargeted, or may cross the boundary between the system of interest and the setting in which that system is embedded).  An example of the use of aggregate senders and receivers is shown in Section~\ref{sec_class}.  A single action or relational event, $a$, is then defined to be a tuple containing the sender of that action $s=s(a)\in S$, the receiver of the action $r=r(a)\in R$, the type of action $c=c(a)\in C$, and the time that the action occurred $\tau=\tau(a)$; formally, $a=(s,r,c,t)$, the analog of an edge in a dynamic network setting.  In practice, we may associate one or more covariates with each potential action ($X_a$), relating to properties of the sender or receiver, the sender/receiver dyad, the time period in question, et cetera.  A series of relational events observed from time 0 (defined to be the onset of observation) and a certain time $t$ comprise an event history, denoted $A_t=\{a_i:\tau(a_i)\le t\}$.  Typically, we will observe a realization of $A_t$ and seek to infer the mechanisms that generated (which will be expressed via a set of parameters, $\theta$, as described below).  At any given point in the event history, the set of possible events (or \emph{support}) is defined by the set $\mathbb{A}(A_t)\subseteq S \times R \times C$, where $\times$ indicates the Cartesian product.  We note that the support may be endogenous, allowing us to consider cases in which particular actions within the event history either make new actions possible or render previously available actions impossible, or exogenous whereby certain possibilities in the support have been restricted (or otherwise new opportunities availed) due to circumstances outside of the system under study.  (For instance, an individual who has left a building cannot speak to those still within it, and the appearance of a new entrant provides a new target for interaction.)

Let $\mathbb{A}$ define the set of events that are possible at any moment.  The propensity of such an event to occur is defined via its \emph{hazard}, i.e. the limit of the conditional rate of event occurrence in a time window about a given point, as the width of that window approaches 0.  Intuitively, the hazard of relational event $a$ at time $t$ is non-negative and equal to 0 if and only if $a\not\in\mathbb{A}(A_t)$ (i.e., $a$ is currently impossible); larger hazards correspond to higher propensities.   It is important to note that each event that is possible at a given moment has a non-zero hazard, and not merely those events that happen to occur; by observing both the events that transpired and the events that could have transpired (but did not), we seek to infer the propensities for all possible events.  Such inference requires that we parameterize our event hazards, and it is natural to conceive of each as arising from a combination of mechanistic factors that either enhance or inhibit the realization of the event in question.  Typically, we implement this by asserting that the hazard of each event is a multiplicative function of a series of statistics, each of which encodes the effect of a given mechanism on event propensity.  Formally, this is expressed as:
\begin{equation*}
\lambda_{a A_t \theta} = \begin{cases} \exp\left(\theta^T u\left(s(a),r(a),c(a),X_a,A_t\right)  \right) & \text{if } a\in \mathbb{A}(A_t)\\ 0 & \text{otherwise} \end{cases},
\end{equation*}
where $\lambda_{a A_t \theta}$ is the hazard of potential event $a$ at time $t$ given history $A_t$, $\theta$ is a vector of real-valued parameters, and $u$ is a vector of functions (i.e., statistics) that may depend upon the sender, receiver, or type of an event, covariates, and/or the prior event history.  It should be noted that the log-linear form for the hazard function used above is not strictly necessary, and other forms are possible. However, we do not consider such alternatives here.

The role played by the $u$ functions in a relational event model is analogous to that of the sufficient statistics in an exponential random graph model \citep[see, e.g.][]{wasserman.robins:ch:2005}, or to the effects in a conventional hazard model \citep{blossfeld.rohwer:bk:1995}: each represents a mechanism that may increase or decrease the propensity of a given action to be taken, as governed by $\theta$.  Each unit change in $u_i$ multiplies the hazard of an associated event by $\exp(\theta_i)$, thereby making it (ceteris paribus) more prevalent and quick to occur or less prevalent and slower to occur.  Typically, candidates for $u$ are proposed on a priori theoretical grounds, with $\theta$ being inferred from available data.  Comparison of goodness-of-fit for models with alternative choices of $u$ allows for alternative theories of social mechanisms to be tested, without assuming that the dynamics are governed by any single mechanism.

\begin{figure}[ht]
\centering
\includegraphics[width=\textwidth]{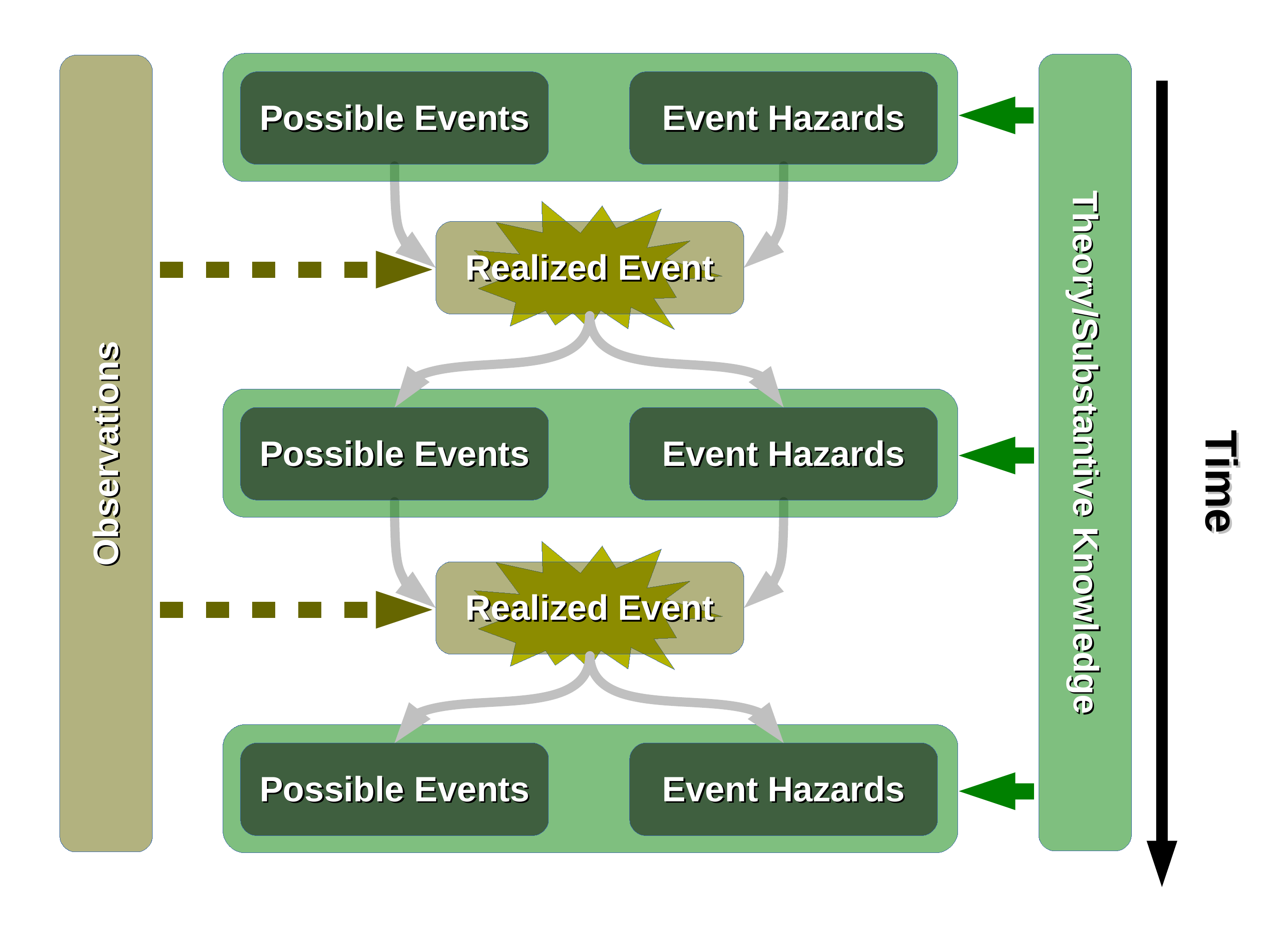}
\caption{Schematic representation of the inferential logic of the relational event framework.  Models, proposed on theoretical grounds, determine the set of possible events and the mechanisms governing event hazards; observations of realized events are employed to infer unknown parameters governing the strengths and directions of effects, and to select among competing models. \label{fig.cartoon}}
\end{figure}

Figure~\ref{fig.cartoon} illustrates the logic of relational event framework by depicting a very general relational event process together with its theoretical components.  In this figure, time runs downward from the top of the illustration to the bottom (as indicated by the rightmost vertical axis). We begin with the state of the world \emph{prior to} any observation of a relational event.  This state can be characterized by the set of potential actions (or possible events) and their underlying propensities to occur (or their respective event hazards).  For example, we may observe a group of individuals in a room, each of whom my direct a speech act at the others, with the hazards representing the distribution of action propensities.  Then, something happens: we observe a realized relational event---one of the actors (the sender) addresses another actor (the receiver).  The occurrence of this particular action, in turn, may have changed the state of the world, possibly including what actions are possible and each individual's propensity to act.  For instance, speaking first may have emboldened the first sender and incremented her propensity to speak even more. Thus, we update the set of possible events and their hazards to reflect new information given the current state of the event history.  Next, something else happens: we observe another relational event.  Again, this event may change the set of possible events and their hazards, and we update our view of the world based on the past history.  This process continues by turns until the last event (not shown).  Just as we make observations on the sequence of events, we use theory and substantive knowledge about the world to make suppositions or impose limits on the set of possible events and to derive the $u$ statistics that govern the event hazards.

As the above indicates, the types of effects we estimate using the relational event framework can capture a wide range of mechanisms involving both endogenous behavioral dynamics and exogenous effects (either covariate-based or the impact of exogenous events).  Typical examples include actor-level fixed effects (rates for sending and receiving events for each actor), sub-sequence effects, and time invariant and time variant covariate effects.  There are many possibilities for modeling endogenous dynamics using the relational event framework because there are many types of event history sub-sequences from which one may build sufficient statistics.  Some such sub-sequences are of general theoretical interest.  For example, we may consider the social processes related to the persistence of action, order of action, exchanges within triads of actors, conversational dynamics, or even dynamic preferential attachment.  Each of these processes can be parameterized in terms of a series of prior events in the life history, allowing it to be implemented in the relational event framework.  The selection of such effects to proposed in a candidate model should be driven by the research question and evaluated by assessing goodness-of-fit (options currently supported by software are listed in the tutorial, below).  For example, much research has shown that persons who have interacted frequently in the past are likely to continue to interact in the future.  In a relational event context, we might thus hypothesize that sending events to certain individuals increases the chances that they will remain the targets of events in the future.  This behavior may be characterized as a type of social persistence or inertia and can be implemented with an effect that treats the fraction of previous contacts as a predictor of future contact.  We might also hypothesize that the order in which one received ties from others in the past plays a role in one's likelihood of replying.  Specifically, because the last thing that happened is very likely to be the most salient, we may model this process with a statistic that employs the inverse of the order of an actor's receipt of events from others as a predictor of that actor's sending of events back to them in the future.  If the inclusion of this effect in the model substantially improves fit (net of degrees of freedom consumed), we conclude that the mechanism in question is predictive of the observed social process; if, however, we do not find such an improvement, we may thereby conclude that the observed pattern of interaction does not support the presence of the proposed mechanism.  We return to more examples of relational event effects in the tutorial, below.  

Regardless of which behaviors (or covariates) are of interest, it is important to understand the basic assumptions of the model used to estimate their effects on the relational event process; further details can be found in \citet{butts2008rems}.  Here, we briefly review three of the most relevant assumptions that most modelers should understand before using the relational event framework.  First, we assume that all events are recorded, and that the onset of the observation period is exogenously determined (e.g., chosen by the researcher or set by a random external event).  Second, we assume that no events can occur at exactly the same time but, rather, are temporally ordered. This assumption is perhaps the key distinction that separates the relational event regime from the dynamic network regime (as discussed above).  Finally, we typically assume that event hazards and the support are piecewise constant, with changes occurring either when an endogenous event is realized or at exogenous ``clock'' events. This final assumption has numerous useful implications, among them being ease of computation and interpretation, the ability to infer parameters when exact times are unknown, and the fact that the waiting times between events are conditionally exponentially distributed.  (Piecewise constancy is also a standard assumption in the well-known Cox proportional hazards models \citep{mills2011iseh}, where it yields similar advantages.)  While this last assumption can be relaxed, current software implementations of the relational event framework \citep[e.g. the \texttt{relevent} package for \texttt{R}][]{butts2010rrem} employ it.  

Of these assumptions, the most critical is the notion that events are well-ordered in time.  While non-simultaneity is in practice vital only for events whose occurrence can affect each others' hazards, and while there are various workarounds for data sets with small amounts of simultaneity (e.g., due to imprecise coding event times), large numbers of simultaneous events suggest a system which is not in the relational event regime.  Such cases may be better represented as dynamic networks, in the manner discussed above.

While the relational event paradigm is defined in terms of instantaneous events that unfold in continuous time, inference for relational event models does not necessarily require that event times be known.  It is useful in this regard to distinguish two general cases: event histories in which only the order of events is known (``ordinal time''); and event histories in which the exact time between events is known (``exact'' or ``interval time'').  \citet{butts2008rems} derives the model likelihood for both scenarios under the assumptions listed above.  Importantly, under the assumption of piecewise constant hazards, the parameter vector $\theta$ can in principle be identified up to a pacing constant; since relative rather than absolute hazards are typically of primary scientific interest, this implies that information on event ordering is frequently adequate to employ the framework.  Such data is common e.g. in archival or observational settings, in which it may be feasible to construct a transcript of actions taken but difficult or impossible to time them accurately.  Both the ordinal and exact cases can be analyzed using the \texttt{relevent} package which, supports a variety of model effects.  Additionally, while we are here focused on the basic case dyadic relational event models in a single event history, the framework is general enough to accommodate multiple event histories and even ego-centered event histories \citep{dubois&2013hmre,marcum&butts2015csse} should one possess those types of data. 

\section{Sample Cases}

To illustrate the use of the relational event model (REM) family, we employ sample case data from two previously published sources. First, to illustrate the relational event model for ordinal time data, we use data from \citet{butts3&2007rcnw}. These data consist of radio communications among 37 named communicants in a police unit that responded to the World Trade Center disaster on the morning of September 11$^{th}$, 2001. Second, to illustrate REMs for exactly timed data, we use a time-modified\footnote{Some events were given in order, but not distinguished by time; these have been spaced by 0.1 min for purposes of illustration.}, subset of data from Dan McFarland who recorded conversations occurring between 20 participants in classroom discussions \citep{bender&mcfarland2006asdn}. Both datasets are available online for didactic purposes here: \href{http://statnet.org}{http://statnet.org}.

For the \texttt{relevent} software package used in the tutorial below, data are stored in ``rectangular" format as an $m \times 3$ matrix we call an ``edgelist" (where $m$ is the number of events).  The first column of the edgelist indexes either the time or the order of the events, depending on the type of data.  The second and third columns index the senders and receivers of the events, respectively, numbered from 1 to $n$ (where $n$ is the number of interacting parties).  Importantly, the edgelist must be ordered by the first column (i.e., by time or event order).  For exact timing data, the last row of the edgelist should index a null event for the time that observation period ended (by default, any event occurring in this row will be ignored by the software). 

Optional sender and receiver covariate data may be stored separate from the edgelist as vectors or arrays, provided that they are ordered consistently with the actor set (1 through the number of actors).  For time invariant covariates, this will be an $n \times p$ matrix, where $n$ indexes the actors and $p$ indexes the covariates. For time varying actor covariates, data should be stored in a 3-dimensional $m \times p \times n$ array, where $m$ indexes time and $p$ and $n$ index covariates and actors as above. 

Optional event covariate data may be stored similarly.  For time invariant covariates, the data should be stored in a 3-dimensional $p \times n \times n$ array, where $p$ and $n$ index each fixed covariate and actor, respectively. Likewise, time varying event covariates should be stored in a 4-dimensional $m \times p \times n \times n$ array, where $m$ indexes time and the other dimensions are as above. 

\subsection{Butts et al.'s WTC Data}
The 9/11 terrorist attacks at the World Trade Center (WTC) in New York City in 2001 set off a massive response effort, with police being among the most prominent.  As in much routine police work, radio communication was essential in coordinating activities during the crisis. \citet{butts3&2007rcnw} coded radio communication events between officers responding to 9/11 from transcripts of communications recorded during the event.  We will illustrate ordinal time REMs using the 481 communication events from 37 named communicants in that data set.  It is important to note that the WTC radio data was coded from transcripts that lacked detailed timing information; we do not therefore know precisely when these calls were made.  We do, however, know the order in which calls were made, and can use this to fit temporally ordinal relational event models. Additionally, we will employ a single actor-level covariate from this dataset: an indicator for whether or not a communicant filled an institutional coordinator role, such as a dispatcher \citep{prahova&butts2008ecwt}.

\subsection{McFarland's Classroom Data}

Dan McFarland's classroom dataset includes exactly timed interactions between students and instructors within a high school classroom  \citep{mcfarland2001srhf,bender&mcfarland2006asdn}. Sender and receiver communication events (n=691) were recorded between 20 actors (18 students and 2 teachers) along with the time of the events in increments of minutes.  The data employed here were modified slightly to increase the amount of time occurring between very closely recorded events, ensuring no simultaneity of events as assumed by the relational event framework.  Two actor-level covariates are also at hand in the dataset used here: whether the actor was a teacher and whether the actor was female.

\section{Tutorial}
Software for fitting relational event models is provided by the \texttt{relevent} package for \texttt{R} \citep{butts2010rrem}. There are numerous tutorials available online that provide instruction on how to obtain and learn to use the free \texttt{R} software.  We direct neophyte users to the R project website (CRAN) to browse those resources: \href{https://cran.r-project.org/}{https://cran.r-project.org/}.  In this tutorial we assume that \texttt{R} is installed and users have some experience with statistical programming in that environment. 

The \texttt{relevent} package and it's dependencies can be downloaded from CRAN using \texttt{R}, installed, and loaded into the user's environment in the usual manner:

\begin{lstlisting}
> install.packages("relevent")
> library(relevent)
\end{lstlisting}

Assuming that the data object was downloaded to one's working directory, it can likewise be loaded in the usual manner using \texttt{load(``remdata.Rdata")} (assuming one has downloaded the data from the URL given above and renamed it thusly).  The R-data object contains 11 objects, which will be used and described throughout this tutorial:
\begin{lstlisting}
> load("remdata.Rdata")
> ls()
 [1] "as.sociomatrix.eventlist" "Class"                   
 [3] "ClassIntercept"           "ClassIsFemale"           
 [5] "ClassIsTeacher"           "sleep.glbs"              
 [7] "sleep.int"                "wtc.coord"               
 [9] "WTCPoliceCalls"           "WTCPoliceIsICR"          
[11] "WTCPoliceNet"
\end{lstlisting}

Dyadic relational event models are intended to capture the behavior of systems in which individual social units (persons, organizations, animals, etc.) direct discrete actions towards other individuals in their environment. Within the \texttt{relevent} package, the \texttt{rem.dyad()} function is the primary workhorse for modeling dyadic data.  From the supplied documentation in \texttt{R}, the \texttt{rem.dyad()} function definition lists a number of arguments and parameters:

\begin{verbatim}
rem.dyad(edgelist, n, effects = NULL, ordinal = TRUE, acl = NULL,
       cumideg = NULL, cumodeg = NULL, rrl = NULL, covar = NULL, ps = NULL,
       tri = NULL, optim.method = "BFGS", optim.control = list(), 
       coef.seed = NULL, hessian = FALSE, sample.size = Inf, verbose = TRUE, 
       fit.method = c("BPM", "MLE", "BSIR"), conditioned.obs = 0, 
       prior.mean = 0, prior.scale = 100, prior.nu = 4, sir.draws = 500, 
       sir.expand = 10, sir.nu = 4, gof = TRUE)
\end{verbatim}
In this tutorial, we focus on the first four arguments---\emph{edgelist, n, effects, ordinal}; the ninth argument \emph{covar}; and the fifteenth argument \emph{hessian}.  The remaining arguments govern model fitting procedures and output and their default values will suffice here.  The first argument, \emph{edgelist}, is how the user passes their data to \texttt{rem.dyad}; aptly, this takes an edgelist as described above.  The second argument, $n$, should be a single integer representing the number of actors in the network.  The third argument, \emph{effects}, is how the user specifies which statistics (effects) will be used to model the data.  This argument should be a character vector where each element is one or more of the following pre-defined effect names:

\clearpage

{\singlespace
\begin{compactitem}
\item `NIDSnd': Normalized indegree of v affects v's future sending rate
\item `NIDRec': Normalized indegree of v affects v's future receiving rate
\item `NODSnd': Normalized outdegree of v affects v's future sending rate
\item `NODRec': Normalized outdegree of v affects v's future receiving rate
\item `NTDegSnd': Normalized total degree of v affects v's future sending rate
\item `NTDegRec': Normalized total degree of v affects v's future receiving rate
\item `FrPSndSnd': Fraction of v's past actions directed to v' affects v's future rate of sending to v'
\item `FrRecSnd': Fraction of v's past receipt of actions from v' affects v's future rate of sending to v'
\item `RRecSnd': Recency of receipt of actions from v' affects v's future rate of sending to v'
\item `RSndSnd': Recency of sending to v' affects v's future rate of sending to v'
\item `CovSnd': Covariate effect for outgoing actions (requires a `covar' entry of the same name)
\item `CovRec': Covariate effect for incoming actions (requires a `covar' entry of the same name)
\item `CovInt': Covariate effect for both outgoing and incoming actions (requires a `covar' entry of the same name)
\item `CovEvent': Covariate effect for each (v,v') action (requires a `covar' entry of the same name)
\item `OTPSnd': Number of outbound two-paths from v to v' affects v's future rate of sending to v'
\item `ITPSnd': Number of incoming two-paths from v' to v affects v's future rate of sending to v'
\item `OSPSnd': Number of outbound shared partners for v and v' affects v's future rate of sending to v'
\item `ISPSnd': Number of inbound shared partners for v and v' affects v's future rate of sending to v'
\item `FESnd': Fixed effects for outgoing actions
\item `FERec': Fixed effects for incoming actions
\item `FEInt': Fixed effects for both outgoing and incoming actions
\item `PSAB-BA': P-Shift effect (turn receiving) - AB$\rightarrow$BA (dyadic)
\item `PSAB-B0': P-Shift effect (turn receiving) - AB$\rightarrow$B0 (non-dyadic)
\item `PSAB-BY': P-Shift effect (turn receiving) - AB$\rightarrow$BY (dyadic)
\item `PSA0-X0': P-Shift effect (turn claiming) - A0$\rightarrow$X0 (non-dyadic)
\item `PSA0-XA': P-Shift effect (turn claiming) - A0$\rightarrow$XA (non-dyadic)
\item `PSA0-XY': P-Shift effect (turn claiming) - A0$\rightarrow$XY (non-dyadic)
\item `PSAB-X0': P-Shift effect (turn usurping) - AB$\rightarrow$X0 (non-dyadic)
\item `PSAB-XA': P-Shift effect (turn usurping) - AB$\rightarrow$XA (dyadic)
\item `PSAB-XB': P-Shift effect (turn usurping) - AB$\rightarrow$XB (dyadic)
\item `PSAB-XY': P-Shift effect (turn usurping) - AB$\rightarrow$XY (dyadic)
\item `PSA0-AY': P-Shift effect (turn continuing) - A0$\rightarrow$AY (non-dyadic)
\item `PSAB-A0': P-Shift effect (turn continuing) - AB$\rightarrow$A0 (non-dyadic)
\item `PSAB-AY': P-Shift effect (turn continuing) - AB$\rightarrow$AY (dyadic)
\end{compactitem}}

The fourth argument, \emph{ordinal}, is a logical indicator that determines whether to use the ordinal or exact timing likelihood. The default setting specifies ordinal timing (TRUE). The ninth argument, \emph{covar}, is how the user passes covariate data to \texttt{rem.dyad()}. Objects passed to this argument should take the form of an \texttt{R} list, where each element of the list is a covariate as described above.  When covariates are indicated, then there should be an associated covariate effect listed in the \emph{effects} argument and each element of the \emph{covar} list should be given the same name as its corresponding effect type specified in \emph{effects} (e.g., `CovSnd', `CovRec', etc). Finally, the fifteenth argument \emph{hessian} is a logical indicator specifying whether or not to compute the Hessian of the log-likelihood or posterior surface, which is used in calculating inferential statistics. The default value of this argument is FALSE. 

Having introduced the relational event package and the model fitting function, we now transition to examples of fitting relational event models using the two datasets described above.  Since the case of ordinal timing is somewhat simpler than that of exact timing, we consider the World Trade Center data first in the tutorial. 

\subsection{Ordinal Time Event Histories}
Before we move to the analysis of the WTC relational event dataset,  it is useful to visually inspect both the raw data and the time-aggregated network. The eventlist is stored in an object called \texttt{WTCPoliceCalls}.  Examining the first six rows of this data reveals that the data is a matrix with the timing information, source (i.e., the sender, numbered from 1 to 37), and recipient (i.e., the receiver, again numbered from 1 to 37) for each event (i.e., radio call):

\clearpage
\begin{lstlisting}
> head(WTCPoliceCalls)
  number source recipient
1      1     16        32
2      2     32        16
3      3     16        32
4      4     16        32
5      5     11        32
6      6     11        32
\end{lstlisting}
Thus, we can already begin to see the unfolding of a relational event process just by inspecting these data visually. First, we see that responding officer 16 called officer 32 in the first event, officer 32 then called 16 back in the second (which might be characterized as a local reciprocity effect or $AB \rightarrow BA$ participation shift \citep{gibson:sf:2003}. This was followed by 32 being the target of the next four calls, perhaps due to either some unobserved coordinator role that 32 fills in the communication structure or due to the presence of a recency mechanism.  Further visual inspection is certainly warranted here. We can use the included \texttt{sna} function \texttt{as.sociomatrix.edgelist()} to convert the eventlist into a valued sociomatrix, which we can then plot using \texttt{gplot()}:

\begin{lstlisting}
> WTCPoliceNet <- as.sociomatrix.eventlist(WTCPoliceCalls,37)
> gplot(WTCPoliceNet,edge.lwd=WTCPoliceNet^0.75,arrowhead.cex=log(as.edgelist.sna(WTCPoliceNet)[,3])+.25,vertex.col=ifelse(WTCPoliceIsICR,"black","gray"),vertex.cex=1.25,vertex.sides=ifelse(WTCPoliceIsICR,4,100),coord=wtc.coord)
\end{lstlisting}

\begin{figure}
\caption{Time-Aggregated WTC Police Radio Communication Network\label{fig.wtc}}
\centering
\includegraphics[width=\textwidth]{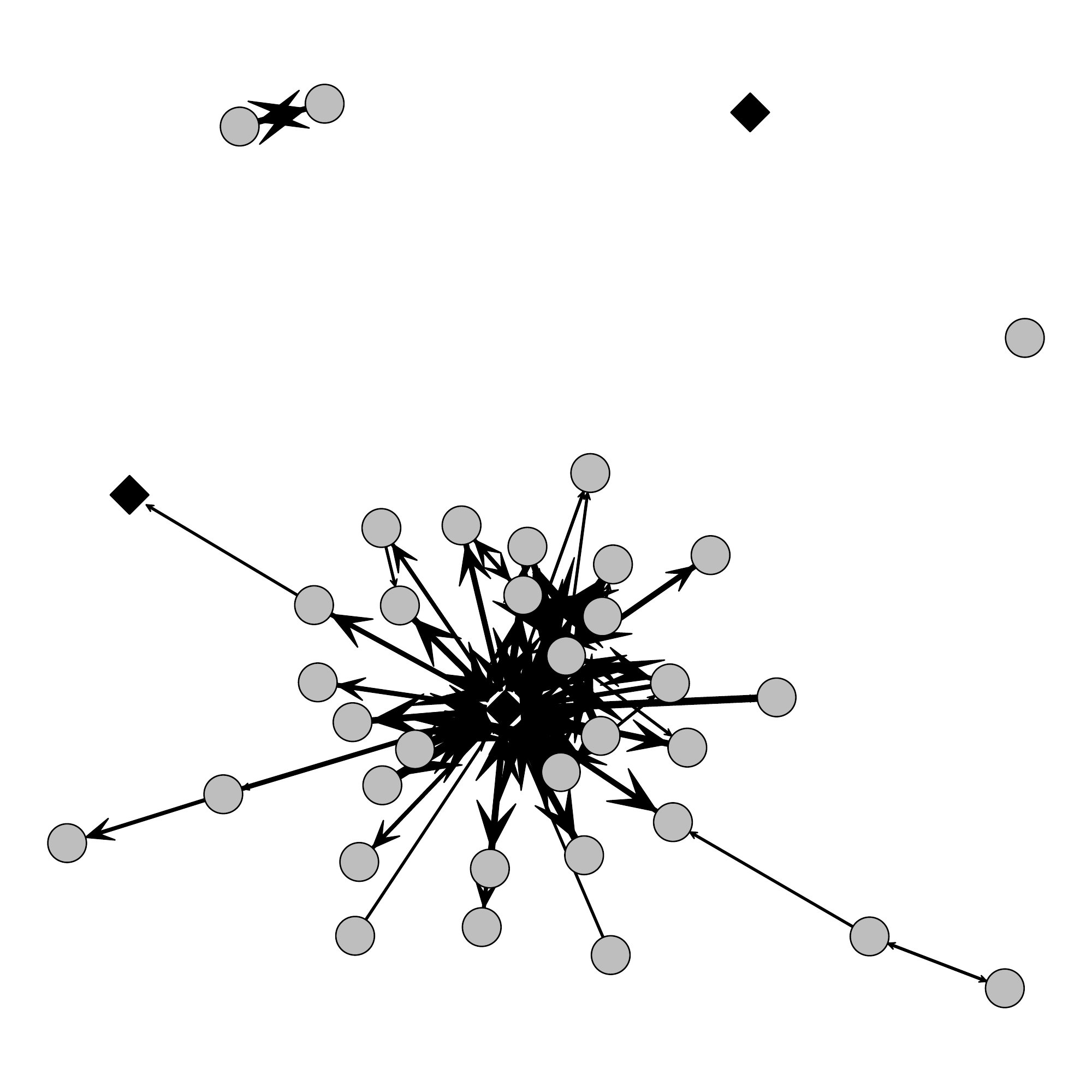}
\end{figure}

Figure~\ref{fig.wtc} is the resulting plot of the time-aggregated WTC Police communication network.  Your own may look slightly different due to both random node placement that \texttt{gplot()} uses to initiate the plot and because  this figure has been tuned for printing.  The three black square nodes represent actors who fill institutional coordinator roles and gray circle nodes represent all other communicants.  A directed edge is drawn between two actors, $i$ and $j$, if actor $i$ ever called actor $j$ on the radio.  The edges and arrowheads are scaled in proportion to the number of calls over time.  There are 37 actors in this network and the 481 communication events have been aggregated to 85 frequency weighted edges.  This is clearly a hub-dominated network with two actors sitting on the majority of paths between all other actors.  While the actor with the plurality of communication ties is an institutional coordinator (the square node at the center of the figure), heterogeneity in sending and receiving communication ties is evident, with several high-degree non-coordinators and two low-degree institutional coordinators, in the network.  This source of heterogeneity is a good starting place from which to build our model. 

\subsubsection{A first model: exploring ICR effects}
We begin by fitting a very simple covariate model, in which the propensity of individuals to send and receive calls depends on whether they occupy institutionalized coordinator roles (ICR). We fit the model by passing the appropriate arguments to \texttt{rem.dyad} and summarize the model fit using the \texttt{summary()} function on the fitted relational event model object. 
\begin{lstlisting}
> wtcfit1<- rem.dyad(WTCPoliceCalls,n=37,effects=c("CovInt"),covar=list(CovInt=WTCPoliceIsICR),hessian=TRUE)
Computing preliminary statistics
Fitting model

Obtaining goodness-of-fit statistics
> summary(wtcfit1)
Relational Event Model (Ordinal Likelihood)

         Estimate  Std.Err Z value  Pr(>|z|)    
CovInt.1 2.104464 0.069817  30.142 < 2.2e-16 ***
---
Signif. codes:  0 '***' 0.001 '**' 0.01 '*' 0.05 '.' 0.1 ' ' 1 
Null deviance: 6921.048 on 481 degrees of freedom
Residual deviance: 6193.998 on 480 degrees of freedom
	Chi-square: 727.0499 on 1 degrees of freedom, asymptotic p-value 0 
AIC: 6195.998 AICC: 6196.007 BIC: 6200.174
\end{lstlisting}

The output gives us the covariate effect, as well as some uncertainty and goodness-of-fit information.  The format is much like the output for a regression model summary, but coefficients should be interpreted per the relational event framework.  In particular, the ICR role coefficient is the logged multiplier for the hazard of an event involving an ICR versus a non-ICR event ( $e^{\lambda_1}$).  This effect is cumulative: an event in which one actor in an ICR calls another actor in an ICR gets twice the log increment ($e^{2\lambda_1}$).  We can see this impact in real terms as follows, respectively:

\begin{lstlisting}
> #Relative hazard for a non-ICR/ICR vs. a non-ICR/non-ICR event
> exp(wtcfit1$coef)
CovInt.1 
8.202707
> #Relative hazard for an ICR/ICR vs. a non-ICR/non-ICR event (twice the effect)
> exp(2*wtcfit1$coef)
CovInt.1 
67.28441
\end{lstlisting}

In this model, ICR status was treated as a homogenous effect on sending and receiving.  Next, we evaluate whether it is worth treating these effects separately with respect to ICR status.  To do so, we enter the ICR covariate as a sender and receiver covariate, respectively, and then evaluate which model is preferred by BIC (lower is better):

\clearpage

\begin{lstlisting}
> wtcfit2<-rem.dyad(WTCPoliceCalls,n=37,effects=c("CovSnd","CovRec"),covar=list(CovSnd=WTCPoliceIsICR,CovRec=WTCPoliceIsICR),hessian=TRUE)
Computing preliminary statistics
Fitting model

> summary(wtcfit2)
summary(wtcfit2)
Relational Event Model (Ordinal Likelihood)

         Estimate  Std.Err Z value  Pr(>|z|)    
CovSnd.1 1.979175 0.095745  20.671 < 2.2e-16 ***
CovRec.1 2.225720 0.092862  23.968 < 2.2e-16 ***
---
Signif. codes:  0 '***' 0.001 '**' 0.01 '*' 0.05 '.' 0.1 ' ' 1 
Null deviance: 6921.048 on 481 degrees of freedom
Residual deviance: 6190.175 on 479 degrees of freedom
	Chi-square: 730.8731 on 2 degrees of freedom, asymptotic p-value 0 
AIC: 6194.175 AICC: 6194.2 BIC: 6202.527

> wtcfit1$BIC-wtcfit2$BIC
[1] -2.352663
\end{lstlisting}
While there appear to be significant ICR sender and receiver effects, their differences do not appear to be large enough to warrant the more complex model (as indicated by the slightly smaller Bayesian Information Criterion (BIC) of the first model).  Smaller deviance-based information criteria, such as the BIC, indicate better model fit. 

\subsubsection{Bringing in endogenous social dynamics}

One of the attractions of the relational event framework is its ability to capture endogenous social dynamics.  Next, we examine several  mechanisms that could conceivably impact communication among participants in the WTC police network.  In each case, we first fit a candidate model, then compare that model to our best fitting model thus far identified. Where effects result in an improvement (as judged by the BIC), we include them in subsequent models, just as we decided for the comparison of the ICR covariate models.

To begin, we note that this is radio communication data.  Radio communication is governed by strong conversational norms (in particular, radio standard operating procedures), which among other things mandate systematic turn-taking reciprocity.  We can test for this via the use of what \citet{gibson2003psod} calls ``participation shifts."  In particular, the \emph{AB-BA} shift, which captures the tendency for B to call A, given that A has just called B, is likely at play in radio communication. Statistics for these effects are described above.  Building from our first preferred model, we now add this dynamic reciprocity term by including \emph{``PSAB-BA"} in the \emph{effects} argument to \texttt{rem.dyad()}:

\begin{lstlisting}
> wtcfit3<-rem.dyad(WTCPoliceCalls,n=37,effects=c("CovInt","PSAB-BA"),covar=list(CovInt=WTCPoliceIsICR),hessian=TRUE)
Computing preliminary statistics

Fitting model
Obtaining goodness-of-fit statistics

> summary(wtcfit3) 
Relational Event Model (Ordinal Likelihood)

         Estimate Std.Err Z value  Pr(>|z|)    
CovInt.1  1.60405 0.11500  13.949 < 2.2e-16 ***
PSAB-BA   7.32695 0.10552  69.436 < 2.2e-16 ***
---
Signif. codes:  0 '***' 0.001 '**' 0.01 '*' 0.05 '.' 0.1 ' ' 1 
Null deviance: 6921.048 on 481 degrees of freedom
Residual deviance: 2619.115 on 479 degrees of freedom
	Chi-square: 4301.933 on 2 degrees of freedom, asymptotic p-value 0 
AIC: 2623.115 AICC: 2623.14 BIC: 2631.467 

> wtcfit1$BIC-wtcfit3$BIC
[1] 3568.707
\end{lstlisting}
It appears that there is a very strong reciprocity effect and that the new model is preferred over the simple covariate model. In fact, the ``PSAB-BA'' coefficient indicates reciprocation events have more than 1500 times the hazard of other types of events ($e^{7.32695}=1520.736$) that might terminate the $AB-BX$ sub-sequence.  

Of course, other conversational norms may also be at play in radio communication.  For instance, we may expect that the current participants in a communication are likely to initiate the next call and that one's most recent communications may be the most likely to be returned.  These processes can be captured with the participation shifts for dyadic turn receiving/continuing and recency effects, respectively:

\begin{lstlisting}
> #Model 4 includes p-shift effects
> wtcfit4<-rem.dyad(WTCPoliceCalls,n=37,effects=c("CovInt","PSAB-BA","PSAB-BY","PSAB-AY"),covar=list(CovInt=WTCPoliceIsICR),hessian=TRUE)
Computing preliminary statistics
Fitting model
Obtaining goodness-of-fit statistics

> summary(wtcfit4)
Relational Event Model (Ordinal Likelihood)

         Estimate Std.Err Z value  Pr(>|z|)    
CovInt.1  1.54283 0.11818 13.0549 < 2.2e-16 ***
PSAB-BA   7.49955 0.11418 65.6831 < 2.2e-16 ***
PSAB-BY   1.25941 0.25131  5.0115 5.402e-07 ***
PSAB-AY   0.87215 0.30612  2.8491  0.004384 ** 
---
Signif. codes:  0 '***' 0.001 '**' 0.01 '*' 0.05 '.' 0.1 ' ' 1 
Null deviance: 6921.048 on 481 degrees of freedom
Residual deviance: 2595.135 on 477 degrees of freedom
	Chi-square: 4325.913 on 4 degrees of freedom, asymptotic p-value 0 
AIC: 2603.135 AICC: 2603.219 BIC: 2619.839

> wtcfit3$BIC-wtcfit4$BIC
[1] 12.62806
\end{lstlisting}

\clearpage
\begin{lstlisting}
>#Model 5 adds recency effects to model 4
> wtcfit5<-rem.dyad(WTCPoliceCalls,n=37,effects=c("CovInt","PSAB-BA","PSAB-BY","PSAB-AY","RRecSnd","RSndSnd"),covar=list(CovInt=WTCPoliceIsICR),hessian=TRUE)
Computing preliminary statistics

Fitting model
Obtaining goodness-of-fit statistics

> summary(wtcfit5)
Relational Event Model (Ordinal Likelihood)

         Estimate Std.Err Z value  Pr(>|z|)    
RRecSnd   2.38495 0.27447  8.6892 < 2.2e-16 ***
RSndSnd   1.34623 0.22307  6.0350 1.590e-09 ***
CovInt.1  1.07058 0.14244  7.5160 5.640e-14 ***
PSAB-BA   4.88714 0.15293 31.9569 < 2.2e-16 ***
PSAB-BY   1.67939 0.26116  6.4304 1.273e-10 ***
PSAB-AY   1.39017 0.31057  4.4762 7.597e-06 ***
---
Signif. codes:  0 '***' 0.001 '**' 0.01 '*' 0.05 '.' 0.1 ' ' 1 
Null deviance: 6921.048 on 481 degrees of freedom
Residual deviance: 2308.413 on 475 degrees of freedom
	Chi-square: 4612.635 on 6 degrees of freedom, asymptotic p-value 0 
AIC: 2320.413 AICC: 2320.591 BIC: 2345.469

> wtcfit4$BIC-wtcfit5$BIC
[1] 274.3701
\end{lstlisting}
The results indicate that turn-receiving, turn-continuing, and recency effects are all at play in the relational event process.  Both models improve over the previous iterations by BIC, and the effect size reciprocity as been greatly reduced by controlling for other effects that reciprocity may have been masking in model 5 (i.e., the ``PSAB-BA" coefficient was reduced from $>7$ to $>4$).  

Finally, recall that our inspection the time-aggregated network in Figure~\ref{fig.wtc} revealed a strongly hub-dominated network, with a few actors doing most of the communication.  Could this be explained in part via a preferential attachment mechanism (per \citet{price1976gtbo} and others), in which those having the most air time become the most attractive targets for others to call?  We can investigate this by including normalized total degree as a predictor of tendency to receive calls:

\begin{lstlisting}
> wtcfit6<-rem.dyad(WTCPoliceCalls,n=37,effects=c("CovInt","PSAB-BA","PSAB-BY","PSAB-AY","RRecSnd","RSndSnd","NTDegRec"),covar=list(CovInt=WTCPoliceIsICR),hessian=TRUE)
Computing preliminary statistics
Fitting model
Obtaining goodness-of-fit statistics

> summary(wtcfit6)
Relational Event Model (Ordinal Likelihood)

         Estimate Std.Err Z value  Pr(>|z|)    
NTDegRec  3.13453 0.56678  5.5305 3.194e-08 ***
RRecSnd   2.02903 0.28500  7.1194 1.084e-12 ***
RSndSnd   0.87116 0.23846  3.6533 0.0002589 ***
CovInt.1  0.70734 0.16400  4.3129 1.611e-05 ***
PSAB-BA   5.32576 0.18236 29.2042 < 2.2e-16 ***
PSAB-BY   1.86023 0.26322  7.0673 1.579e-12 ***
PSAB-AY   1.64806 0.31092  5.3005 1.155e-07 ***
---
Signif. codes:  0 '***' 0.001 '**' 0.01 '*' 0.05 '.' 0.1 ' ' 1 
Null deviance: \textsc{6921.048} on 481 degrees of freedom
Residual deviance: \textsc{2277.263} on 474 degrees of freedom
	Chi-square: 4643.785 on 7 degrees of freedom, asymptotic p-value 0 
AIC: 2291.263 AICC: 2291.5 BIC: 2320.494 

> wtcfit5$BIC-wtcfit6$BIC
[1] 24.97434
\end{lstlisting}
Though still significant in the presence of preferential attachment effects, recency and ICR effect coefficients are reduced while participation shift effects are relatively unchanged. This final model is also preferred by BIC and it's clear that the deviance reduction from the null model is quite substantial at 67\%.  While we could continue to investigate additional effects (see the list of options above), model 6 is a good candidate to evaluate model adequacy, which is addressed in the next section.

\subsubsection{Assessing model adequacy}
\begin{figure}
\caption{Histogram of Deviance Residuals from Ordinal Model of WTC Data\label{fig.wtchist}}
\centering
\includegraphics[width=\textwidth]{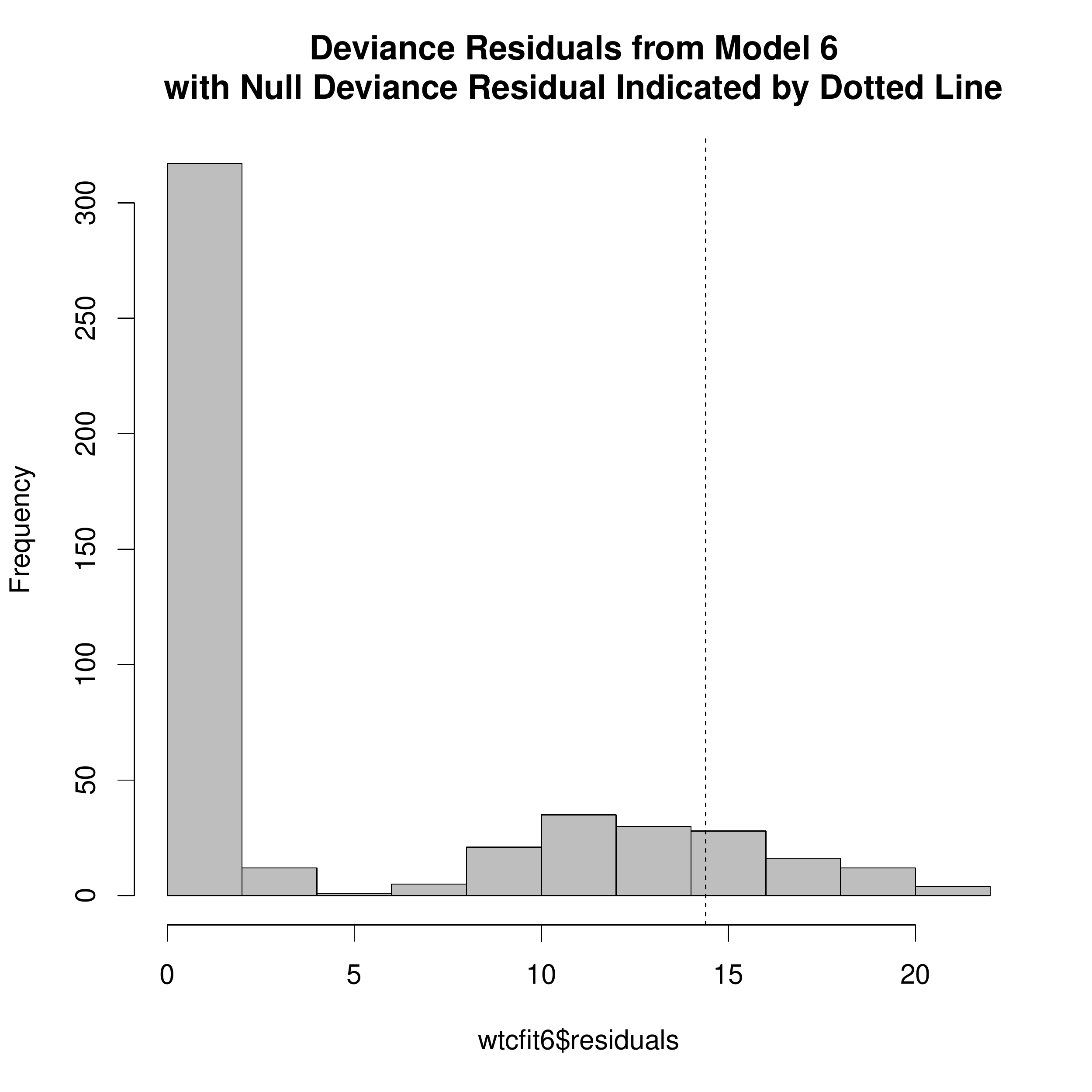}
\end{figure}

Model adequacy is an important consideration: even given that our final model from the exercises above (model 6) is the best of the set, is it good enough for our purposes? There are many ways to assess model adequacy; here, we focus on the ability of the relational event model to predict the next event in the sequence, given those that have come before.  This approach nicely falls within the relational event framework.  A natural question to ask in this framework is how ``surprised" is the model by the data. Put another way, when does the model encounter relational event observations that are relatively poorly predicted?  To investigate this, we can examine the deviance residuals, which are included in the fitted model object.  We begin by calculating the deviance residual under the null which, from the ordinal likelihood derivation in \cite{butts2008rems}, is simply twice the log product of the number of sender-receiver pairs, and comparing that with the deviance residuals under the fitted model: 

\begin{lstlisting}
> #Null deviance residual
> nullresid<- 2*log(37*36)
> #Plot a histogram of the fitted model deviance residuals
> hist(wtcfit6$residuals,main="Deviance Residuals from Model 6 \n with Null Deviance Residual Indicated",col="gray")
> abline(v=nullresid,lty=2)
> #What fraction are below the null resid?
> mean(wtcfit6$residuals<nullresid)
[1] 0.8898129
> #What fraction are less than 3?
> mean(wtcfit6$residuals<3)
[1] 0.6839917
> #How "surprised" is the model?
mean(wtcfit6$residuals>nullresid)
[1] 0.1101871
\end{lstlisting}
The histogram of the model deviance residuals produced from the above code snippet is shown in Figure~\ref{fig.wtchist}.  The dotted line indicates the null deviance residual: the idea here is that we want the model deviance residuals to fall to the right of that cut-off.  Indeed, about 89\% of the model deviance residuals are smaller than the null residual, with 68\% of them being less than 3 (or really, really small).  These initial checks are good conditional evidence that our model is performing really well.  

\clearpage

To investigate further, we can evaluate the extent to which our model could take a random guess about which event comes next and get it right, relative to all possibilities.  Here again, the deviance residuals come in handy as the quantity $e^{\frac{D_i}{2}}$, where $D_i$ is the model deviance residual for event i, is a ``random guessing equivalent".  That is, it is the effective number of events such that a random guess about what happens next would be right as often as expected under the model.  

\begin{lstlisting}
> # Distribution of random guessing equivalents for model 6
> quantile(exp(wtcfit6$residuals/2))
   0%          25%          50%          75%         100% 
1.073634     1.268661     1.739723   204.539040 31632.962350 

> # Distribution of random guessing equivalents for model 1
> quantile(exp(wtcfit1$residuals/2))
       0%       25%       50%       75%      100% 
 390.0003  390.0003  390.0003  390.0003 3199.0591
\end{lstlisting}
At least 50\% of the time our final model needs about 1 in 1.7 guesses to correctly predict the next event.  This is in contrast to our first model with just the intercept term for ICR covariate, which needs about 390 such guesses.  For an overall comparison, consider that the null model would get only 1 out of every 1332 ($36*37$) events correct just guessing at random. 

Model adequacy as measured by surprise can also be visually inspected.  First, one can inspect which events are surprising by adding an indicator for model surprise to the original eventlist:

\begin{lstlisting}
> head(cbind(WTCPoliceCalls,surprise=wtcfit6$residuals>nullresid))
   number source recipient surprise
1       1     16        32    FALSE
2       2     32        16    FALSE
3       3     16        32    FALSE
4       4     16        32    FALSE
5       5     11        32     TRUE
\end{lstlisting}
The code snippet prints just the first five events, but these are enough to get a glimpse into why the model might be surprised.  We can see that the first four events, involving exchanges between actors 16 and 32, are not surprising and appear to involve reciprocity and turn continuing participation shifts. The fifth event, however, is surprising, probably because it involves the sudden interruption of a new caller (actor 11).  Thus, it appears that the model is surprised, perhaps unsurprisingly, when events transpire that are not specified by the model statistics such as third-party effects.

\begin{figure}
\caption{Time-Aggregated `Surprising' Events Network Under the Final Relational Event Model of WTC Radio Communications \label{fig.wtc2}}
\centering
\includegraphics[width=\textwidth]{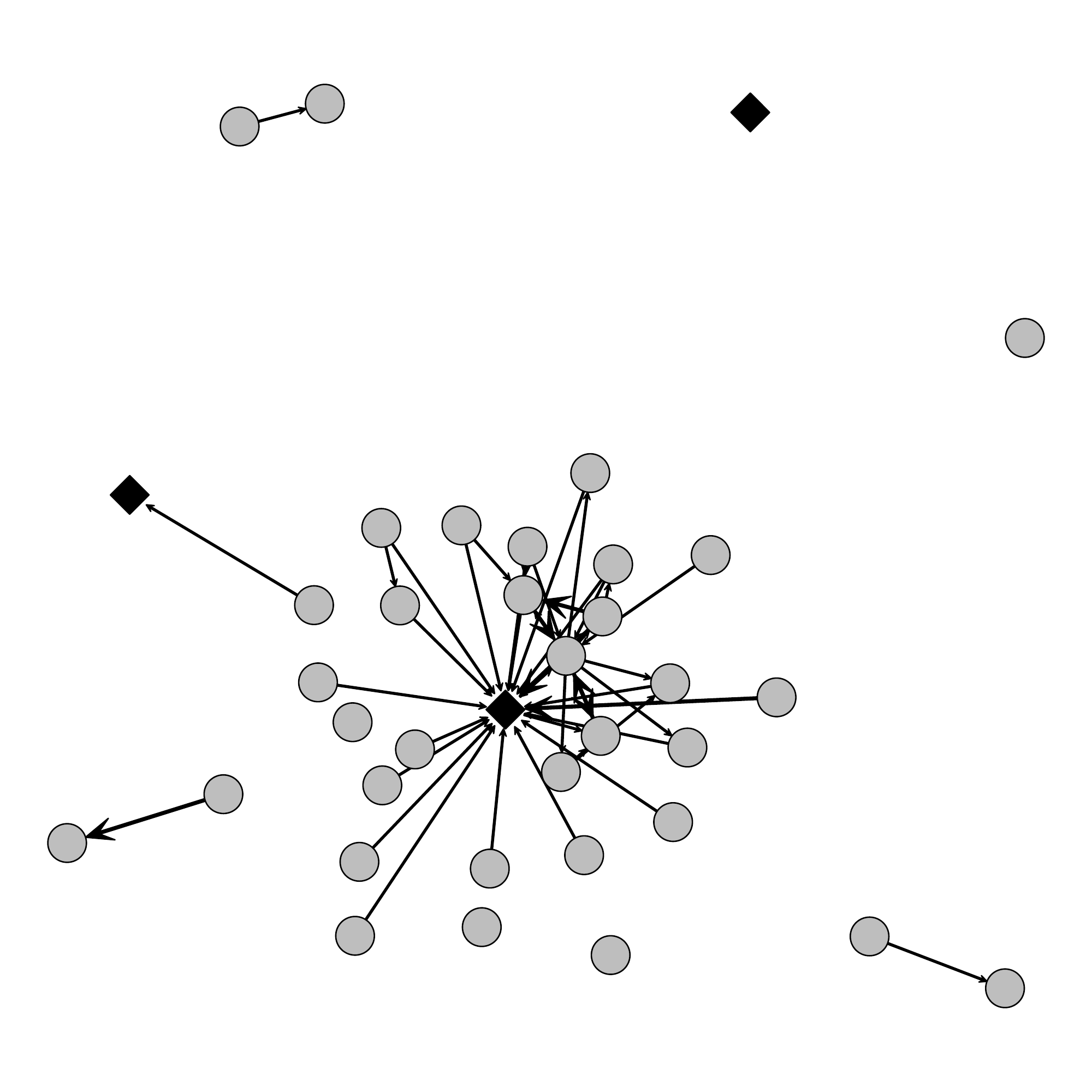}
\end{figure}

These surprising events can also be projected onto the time-aggregated network using \texttt{as.sociomatrix.edgelist}, as before: 

\begin{lstlisting}
>surprising<-as.sociomatrix.eventlist(WTCPoliceCalls[wtcfit6$residuals>nullresid,],37)

>gplot(surprising,edge.lwd=surprising^0.75,arrowhead.cex=log(as.edgelist.sna(surprising)[,3])+.25,vertex.col=ifelse(WTCPoliceIsICR,"black","gray"),vertex.cex=1.25,vertex.sides=ifelse(WTCPoliceIsICR,4,100))
\end{lstlisting}
The resulting plot of the time-aggregated surprising event network is illustrated in Figure~\ref{fig.wtc2}, which can be directly compared with Figure~\ref{fig.wtc}. While there are many fewer events that are surprising than not, it's clear from the figure that the surprising events resolve on where the greatest opportunity for communication exists: namely on calls directed toward the main hub at the center and also calls sent from the secondary hub to others.  This suggests the existence of some unobserved heterogeneity related to those actors not explained by conversational norms, preferential attachment to them, or whether or not they fill institutional coordinator roles. 

Finally, the function \texttt{rem.dyad()} supplies two additional components in returned model objects that are useful for evaluating adequacy. These are the rank of the observed events in the predicted rate structure and an a pair of indicator for whether or not the model exactly predicts the sender and receiver, respectively, involved in each event.  While far more stringent as measures of surprise than the deviance residuals, these statistics can be quite informative for well-fitting models. 

\begin{figure}
\caption{Classification Accuracy of the Observed Ranks Under Model 6 with Prediction Thresholds Indicated at 0.05, 0.1, and 0.25 \label{fig.ecdf}}
\centering
\includegraphics[width=\textwidth]{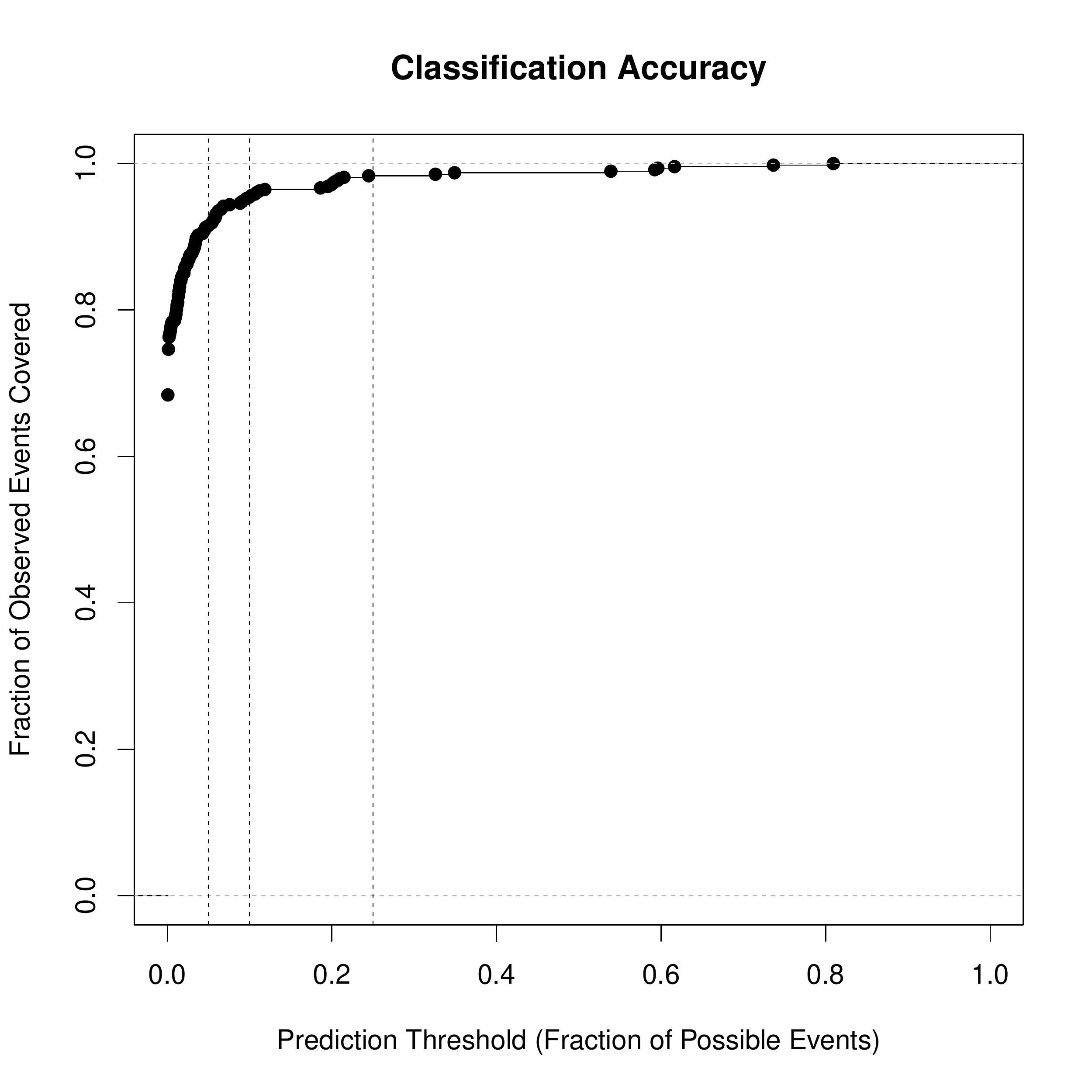}
\end{figure}

For instance, we can inspect the empirical cumulative distribution function of the observed ranks to assess classification accuracy of the model at various thresholds: 
\begin{lstlisting}
> plot(ecdf(wtcfit6$observed.rank/(37*36)),
xlab="Prediction Threshold (Fraction of Possible Events)", ylab="Fraction of Observed Events Covered",main="Classification Accuracy")
> abline(v=c(0.05,0.1,0.25),lty=2)
\end{lstlisting}
The resulting plot of the ECDF is shown in Figure~\ref{fig.ecdf}, which shows that predictions under the model very quickly cover the observed events.  For the strictest measures, we can ask three questions of the exact predictions: 1) what is the fraction of events for which either sender or receiver are exactly predicted; 2) what is the fraction of events for which both sender and receiver are exactly predicted; and, 3) what are the respective fractions of events where we get the sender and receiver right under the model.  These questions are easily addressed using the fitted model output:
\begin{lstlisting}
> mean(apply(wtcfit6$predicted.match,1,any))
[1] 0.7941788

> mean(apply(wtcfit6$predicted.match,1,all))
[1] 0.6839917
 
> colMeans(wtcfit6$predicted.match)
   source recipient 
0.7234927 0.7546778 
\end{lstlisting}

Thus, our final model predicts something right about 79\% of the time (getting the sender right for 72\% and the receiver right about 75\% of the events, respectively) and it predicts the event that actually transpired exactly right 68\% of the time.  Despite its simplicity, this model appears to fit extremely well.  Further improvement is possible, but for many purposes we might view it as an adequate representation of the event dynamics in this WTC police radio communication network.

\subsection{Exact Time Event Histories \label{sec_class}}

We now turn to a consideration of REMs for event histories with exact timing information.  As in the case of ordinal time data, it is useful to begin by examining the raw temporal data and the time-aggregated network.  The edgelist is stored in an object called \texttt{Class}. Printing the first six rows and the last two rows of this object reveals minor differences between the exact time and the ordinal time data structures (discussed above). As before, we have three columns: the event time, the event source (numbered from 1 to 20), and the event target (again, numbered 1 to 20). In this case, event time is given in increments of minutes from onset of observation. Note that the last row of the event list contains the time at which observation was terminated; it (and only it) is allowed to contain NAs, since it has no meaning except to set the period during which events could have occurred. Where exact timing is used, the final entry in the edgelist is always interpreted in this way, and any source/target information on this row is ignored. This row indicates that the total period of observation lasted just over 50 minutes (the length of one class session). 

\begin{lstlisting}
> Class[c(1:6,691:692),]
    StartTime FromId ToId
1       0.135     14   12
2       0.270     12   14
3       0.405     18   12
4       0.540     12   18
5       0.675      1   12
6       0.810     12    1
691    50.910     17    6
692    50.920     NA   NA
\end{lstlisting}
We can again use the sna toolkit to convert and plot the time-aggregated network for inspection. Here, we color the female nodes black and the male nodes gray and represent teachers as square-shaped nodes and students as triangle-shaped nodes. Edges between nodes are likewise scaled proportional to the number of communication events transpiring between actors.
\clearpage

\begin{lstlisting}
 ClassNet<-as.sociomatrix.eventlist(Class,20)
 gplot(ClassNet,vertex.col=ifelse(ClassIsFemale,"black","gray"),vertex.sides=3+ClassIsTeacher,vertex.cex=2, edge.lwd=ClassNet^0.75)
\end{lstlisting}

\begin{figure}
\caption{Time-Aggregated Classroom Communications \label{fig.class}}
\centering
\includegraphics[width=\textwidth]{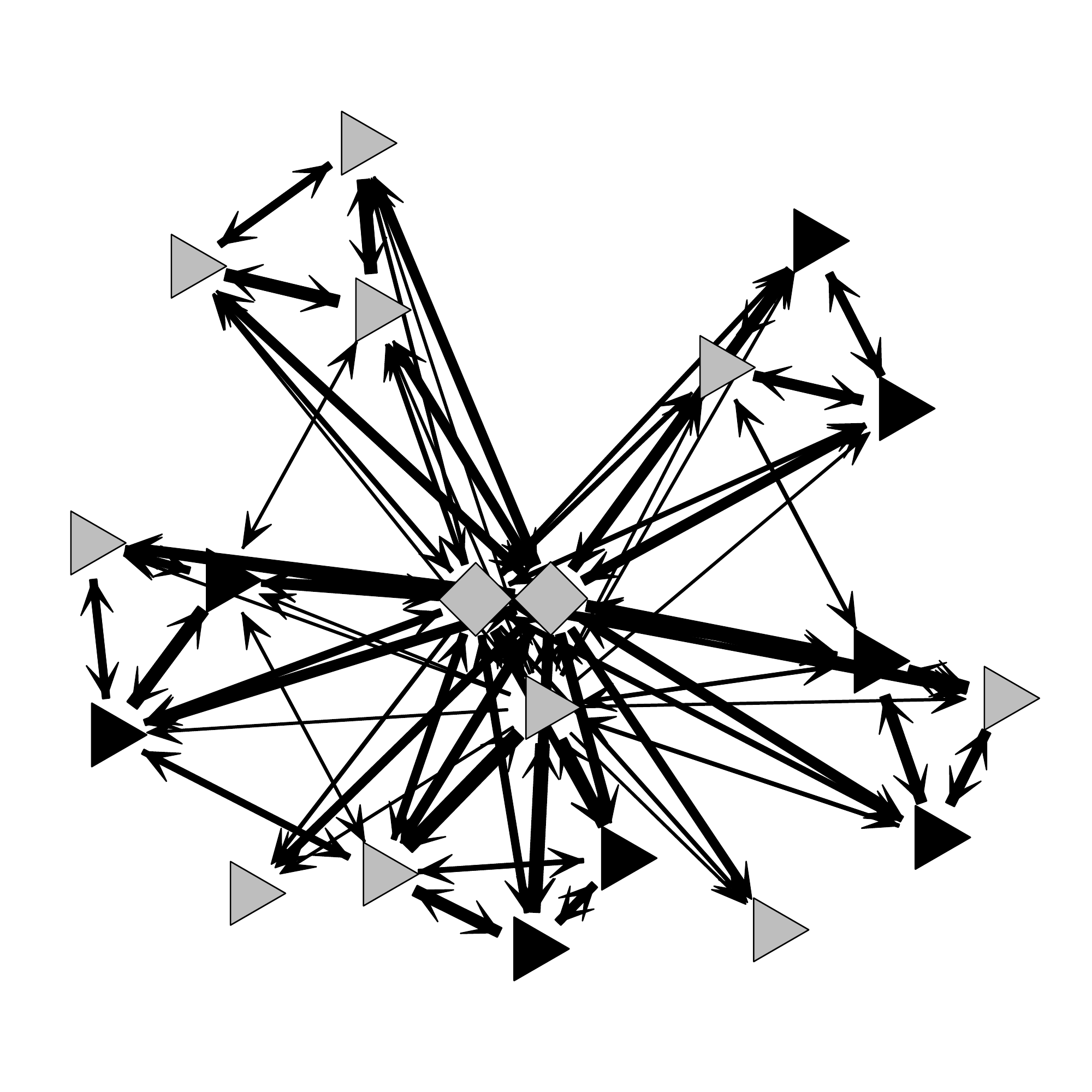}
\end{figure}

Figure~\ref{fig.class} displays the resulting time-aggregated network. A dynamic visualization of this data is also available online in \citep{bender&mcfarland2006asdn} and is well worth examining. While it is clear from this figure that teachers do a great deal of talking, there also appear to be several high-degree students. Female students in this classroom also appear to be slightly more peripheral. Both of these observations warrant inclusion of the respective covariates in our analysis, to which we now turn. 

\clearpage

\subsubsection{Modeling with covariates}
One of the advantages that the exact time relational event model likelihood has over the ordinal time likelihood is its ability to estimate pacing constants (i.e., the global rates at which events transpire).  Here we investigate this with a simple intercept model, containing only a vector of $1s$ as an actor-level sending effect.  This vector is saved as \texttt{ClassIntercept}, which we can pass to the respective covariate arguments in \texttt{rem.dyad()}. Note that we must also tell \texttt{rem.dyad} that we do not want to discard timing information by setting the argument \texttt{ordinal=FALSE}:

\begin{lstlisting}
> classfit1<-rem.dyad(Class,n=20,effects=c("CovSnd"),covar=list(CovSnd=ClassIntercept), ordinal=FALSE,hessian=TRUE)
Computing preliminary statistics
Fitting model

Obtaining goodness-of-fit statistics

> summary(classfit1)
Relational Event Model (Temporal Likelihood)

          Estimate   Std.Err Z value  Pr(>|z|)    
CovSnd.1 -3.332287  0.038042 -87.596 < 2.2e-16 ***
---
Signif. codes:  0 '***' 0.001 '**' 0.01 '*' 0.05 '.' 0.1 ' ' 1 
Null deviance: 5987.221 on 691 degrees of freedom
Residual deviance: 5987.221 on 691 degrees of freedom
	Chi-square: -4.274625e-11 on 0 degrees of freedom, asymptotic p-value 1 
AIC: 5989.221 AICC: 5989.227 BIC: 5993.759 
\end{lstlisting}
The model does not fit any better than the null because it is equivalent to the null model (as indicated by the absence of difference between the null and residual deviance).  As one would expect from first principles, this is really just an exponential waiting time model, calibrated to the observed communication rate.  Thus, to calculate the predicted number of events per minute we may multiply the number of possible event types (here, $20*19=380$) by the coefficient for the intercept:
\clearpage

\begin{lstlisting}
> 380*exp(classfit1$coef)
CovSnd.1 
13.57031 
\end{lstlisting}
This simple model predicts the overall pace of events to occur at nearly 14 events per minute and this matches quite well with the average number of events per minute from the observed data:
\begin{lstlisting}
> (nrow(Class)-1)/max(Class[,1])
[1] 13.57031
\end{lstlisting}.
Because we noted structural heterogeneity based on gender and status in Figure~\ref{fig.class}, we fit a more interesting covariate model that specifies these effects for senders and receivers and evaluate whether there is any improvement over the intercept-only model by BIC. 
\clearpage

\begin{lstlisting}
> classfit2<-rem.dyad(Class,n=20,effects=c("CovSnd","CovRec"), covar=list(CovSnd=cbind(ClassIntercept,ClassIsTeacher,ClassIsFemale), CovRec=cbind(ClassIsTeacher,ClassIsFemale)),ordinal=FALSE,hessian=TRUE)
Computing preliminary statistics
Fitting model

Obtaining goodness-of-fit statistics
> 
> summary(classfit2)
Relational Event Model (Temporal Likelihood)

          Estimate   Std.Err  Z value Pr(>|z|)    
CovSnd.1 -3.834229  0.078842 -48.6319  < 2e-16 ***
CovSnd.2  1.672561  0.091679  18.2436  < 2e-16 ***
CovSnd.3  0.123900  0.094931   1.3052  0.19184    
CovRec.1  0.373733  0.127028   2.9421  0.00326 ** 
CovRec.2  0.165729  0.080896   2.0487  0.04049 *  
---
Signif. codes:  0 '***' 0.001 '**' 0.01 '*' 0.05 '.' 0.1 ' ' 1 
Null deviance: 5987.221 on 691 degrees of freedom
Residual deviance: 5652.318 on 687 degrees of freedom
	Chi-square: 334.9034 on 4 degrees of freedom, asymptotic p-value 0 
AIC: 5662.318 AICC: 5662.405 BIC: 5685.008 
> 
> classfit1$BIC-classfit2$BIC
[1] 308.7508

\end{lstlisting}
With multiple covariates, the model terms ($CovSnd.1$, $CovSnd.2$ etc) are listed in the object in the same order as they were specified within the \texttt{covar} argument. Here, we see a good improvement over the null model but also note that gender does not appear to be predictive of sending communication.  A better model may be one without that specific term included, which we fit below and again compare to the previous model by BIC.
\clearpage
\begin{lstlisting}
> classfit3<-rem.dyad(Class,n=20,effects=c("CovSnd","CovRec"), covar=list(CovSnd=cbind(ClassIntercept,ClassIsTeacher), CovRec=cbind(ClassIsTeacher,ClassIsFemale)),ordinal=FALSE,hessian=TRUE)
Computing preliminary statistics
Fitting model
Obtaining goodness-of-fit statistics

> summary(classfit3)
Relational Event Model (Temporal Likelihood)

          Estimate   Std.Err  Z value  Pr(>|z|)    
CovSnd.1 -3.775227  0.063623 -59.3379 < 2.2e-16 ***
CovSnd.2  1.615762  0.079933  20.2139 < 2.2e-16 ***
CovRec.1  0.371749  0.127020   2.9267  0.003426 ** 
CovRec.2  0.161154  0.080815   1.9941  0.046141 *  
---
Signif. codes:  0 '***' 0.001 '**' 0.01 '*' 0.05 '.' 0.1 ' ' 1 
Null deviance: 5987.221 on 691 degrees of freedom
Residual deviance: 5654.016 on 688 degrees of freedom
	Chi-square: 333.2049 on 3 degrees of freedom, asymptotic p-value 0 
AIC: 5662.016 AICC: 5662.074 BIC: 5680.169 

> classfit2$BIC-classfit3$BIC
[1] 4.839661
\end{lstlisting}
Indeed, there is a marginal improvement in BIC and we retain the model lacking the gender effect for sending communication events.

\subsubsection{Modeling endogenous social dynamics}
While we find that the above covariate models perform better than the null, the final model is still unimpressive in terms of deviance reduction, with only about a 5\% total reduction from the null by our best fitting model.  To investigate further, we propose a set of models that capture endogenous social dynamic effects that are reasonably presumed to be at play in classroom conversations.  These include recency effects and effects that capture aspects of conversational norms, such as turn-taking, sequential address, and turn-usurping.  

As before, we can enter these terms into the model using their appropriate effect names.  We also preserve the covariates from best covariate model (model 3 from the previous section) and check our improvement by BIC. 
\begin{lstlisting}
> #First, just recency effects + model 3:
> classfit4<-rem.dyad(Class,n=20,effects=c("CovSnd","CovRec","RRecSnd","RSndSnd"), covar=list(CovSnd=cbind(ClassIntercept,ClassIsTeacher), CovRec=cbind(ClassIsTeacher,ClassIsFemale)),ordinal=FALSE,hessian=TRUE)
Computing preliminary statistics
Fitting model
Obtaining goodness-of-fit statistics

> #This is preferred:
> classfit3$BIC-classfit4$BIC
[1] 1118.294
 
> #Next conversational norms + model 4 
> classfit5<-rem.dyad(Class,n=20,effects=c("CovSnd","CovRec","RRecSnd","RSndSnd", "PSAB-BA","PSAB-AY","PSAB-BY"),covar=list(CovSnd=cbind(ClassIntercept,ClassIsTeacher), CovRec=cbind(ClassIsTeacher,ClassIsFemale)),ordinal=FALSE,hessian=TRUE)
Computing preliminary statistics
Fitting model
Obtaining goodness-of-fit statistics

> #Again an improvement: 
> classfit4$BIC-classfit5$BIC
[1] 1699.716
\end{lstlisting}
\clearpage

\begin{lstlisting}

> summary(classfit5)
Relational Event Model (Temporal Likelihood)

          Estimate   Std.Err  Z value  Pr(>|z|)    
RRecSnd   2.429233  0.155365  15.6356 < 2.2e-16 ***
RSndSnd  -0.986747  0.144667  -6.8208 9.053e-12 ***
CovSnd.1 -5.003434  0.090609 -55.2201 < 2.2e-16 ***
CovSnd.2  1.253893  0.085160  14.7239 < 2.2e-16 ***
CovRec.1 -0.722690  0.141950  -5.0912 3.559e-07 ***
CovRec.2  0.047936  0.081325   0.5894    0.5556    
PSAB-BA   4.622128  0.137600  33.5910 < 2.2e-16 ***
PSAB-BY   1.677591  0.164930  10.1715 < 2.2e-16 ***
PSAB-AY   2.869968  0.103113  27.8332 < 2.2e-16 ***
---
Signif. codes:  0 '***' 0.001 '**' 0.01 '*' 0.05 '.' 0.1 ' ' 1 
Null deviance: 5987.221 on 691 degrees of freedom
Residual deviance: 2803.315 on 683 degrees of freedom
	Chi-square: 3183.906 on 8 degrees of freedom, asymptotic p-value 0 
AIC: 2821.315 AICC: 2821.58 BIC: 2862.158 
\end{lstlisting}
We can see that adding recency effects to the covariate model results in a much improved fit by BIC. Moreover, there is again an improvement in BIC when conversational norms are added into the model.  The summary of the results from model 5 also show that the remaining gender covariate effect falls out in the presence of the endogenous social dynamic effects.  This hints at the possibility that what seemed at first glance to be a difference in the tendency to receive communication by gender was in fact a result of social dynamics (perhaps stemming from the fact that both instructors are male, with their inherent tendency to communicate more often amplified by local conversational norms). We can confirm that second the gender term is extraneous by evaluating whether a reduced model is preferred by BIC.
\clearpage
\begin{lstlisting}
> classfit6<-rem.dyad(Class,n=20,effects=c("CovSnd","CovRec","RRecSnd","RSndSnd", "PSAB-BA","PSAB-AY","PSAB-BY"),covar=list(CovSnd=cbind(ClassIntercept,ClassIsTeacher), CovRec=ClassIsTeacher),ordinal=FALSE,hessian=TRUE)
Computing preliminary statistics
Fitting model
Obtaining goodness-of-fit statistics
 
> classfit5$AICC-classfit6$AICC
[1] 1.705912
\end{lstlisting}
And, as before, the reduced model is indeed preferred.  We now have a relatively good fitting relational event model specified by a combination of covariate and endogenous dynamic effects.  At this point, we can turn to interpretation of fitted model parameters and model adequacy from our current vantage point. 

\subsubsection{Interpretation of a fitted model}
It is often useful to consider the inter-event times predicted to be observed under various scenarios by a fitted relational event model. Recall that under the piecewise constant hazard assumption, event waiting times are conditionally exponentially distributed.  This allows us to easily work out the consequences of various model effects for social dynamics, at least within the context of a particular scenario.  

The most basic results to interpret from a fitted model are, of course, the coefficients themselves. In interpreting coefficient effects, recall that they act as logged hazard multipliers. Taking their log-inverse (i.e., exponentiating them), produces their hazard multiplier. For instance, the turn-taking participation-shift (p-shift) effect from model 6 has a coefficient value of $4.623682$, which corresponds to an interpretation that response events have about 100 times the hazard of non-response events ($e^{4.623682}=101.8684$).  While this \emph{appears} to be a substantial effect, the fact that an event has an unusually high hazard does not mean that it will necessarily occur.  For instance, while a response of B to a communication from A has hazard that is about 100 times as great as the hazard of a non-$B \rightarrow A$ event all things constant, there are many more events of the latter type.  In fact, there are 379 other events ``competing" with the $B \rightarrow A$ event, and thus the chance that it will occur next is smaller than it may appear by simply taking the hazard multiplier at face value.  This example shows that both relative rates and combinatorics (i.e., the number of possible ways that an event type may occur) govern the result and should temper respective interpretations.

What else can be done with the model coefficients from an interpretation perspective? One basic use of the model coefficients is to examine the expected inter-event times under specific scenarios and conditions. For instance, one may be interested in evaluating the predicted mean inter-event time when nothing else is happening. This is simply governed by the global pacing constant (i.e., the average rate that events transpire, or intercept) and the number of possible events.  Or, one may want to know how long it takes for one actor to respond to another actor given an immediate event (or other such scenarios).  Depending on the model, many of these ``waiting time" effects can be evaluated from coefficients.  To accomplish this using the exact time likelihood, some algebra comes in handy: $\frac{1}{m' \times e^{\sum \lambda}}$, where m is the number of possible events under the scenario and $\lambda$ is the vector of model parameters involving the scenario of interest.  Here again, both the number of ways that an event type can occur ($m'$) and the propensity of such events to occur ($\lambda$) both matter!  

In the following snippet, we evaluate such waiting times under different scenarios from model 6:
\begin{lstlisting}
> #Mean inter-event time if nothing else going on....
> 1/(20*19*exp(classfit6$coef["CovSnd.1"]))
 CovSnd.1 
0.3843285 

> #Mean teacher-student time (again, if nothing else happened)
> 1/(2*18*exp(sum(classfit6$coef[c("CovSnd.1","CovSnd.2")])))
[1] 1.153845
\end{lstlisting}
\clearpage
\begin{lstlisting}

> #Sequential address by teacher w/out prior interaction, given a prior teacher-student interaction, and assuming nothing else happened
> 1/(17*exp(sum(classfit6$coef[c("CovSnd.1","CovSnd.2","PSAB-AY")])))
[1] 0.1384693

> #Teacher responding to a specific student, given an immediate event
> 1/(exp(sum(classfit6$coef[c("CovSnd.1","CovSnd.2","PSAB-BA","RRecSnd")])))
[1] 0.03587346

> #Student responding to a specific teacher, given an immediate event
> 1/(exp(sum(classfit6$coef[c("CovSnd.1","CovRec.1","PSAB-BA","RRecSnd")])))
[1] 0.2657102
\end{lstlisting}
Remember that our temporal units in the classroom dataset are increments of minutes: multiplying these values by 60 returns how many seconds (or fractions thereof) these predicted waiting times entail. Thus, if no other event were to intervene, a teacher would initiate communication with a student after a mean waiting time of approximately 70 seconds.  Given an initial teacher$\to$student communication and no other intervention, the same teacher will produce another speech act after an average of roughly 8 seconds---a rapid-fire lecture mode.  Interestingly, we can also see that teachers are very quick to respond to student communications (a delay of just over 2 seconds, on average), while students take somewhat longer to respond to teachers (about 16 seconds).  Such observations comport well with our general intuition regarding classroom functioning, and illustrate the types of quantitative information that can be gleaned from a REM fit. 

\subsubsection{Assessing model adequacy}
We can assess model adequacy for exact time relational event models in much the same manner as we do for ordinal time models.  The major difference is that we cannot here use a fixed null residual or guessing equivalent.  However, we can still examine ``surprise" based on the deviance residuals of fitted models. 

\begin{figure}
\caption{Histogram of Deviance Residuals from Exact Time Model of McFarland's Classroom Data\label{fig.classhist}}
\centering
\includegraphics[width=\textwidth]{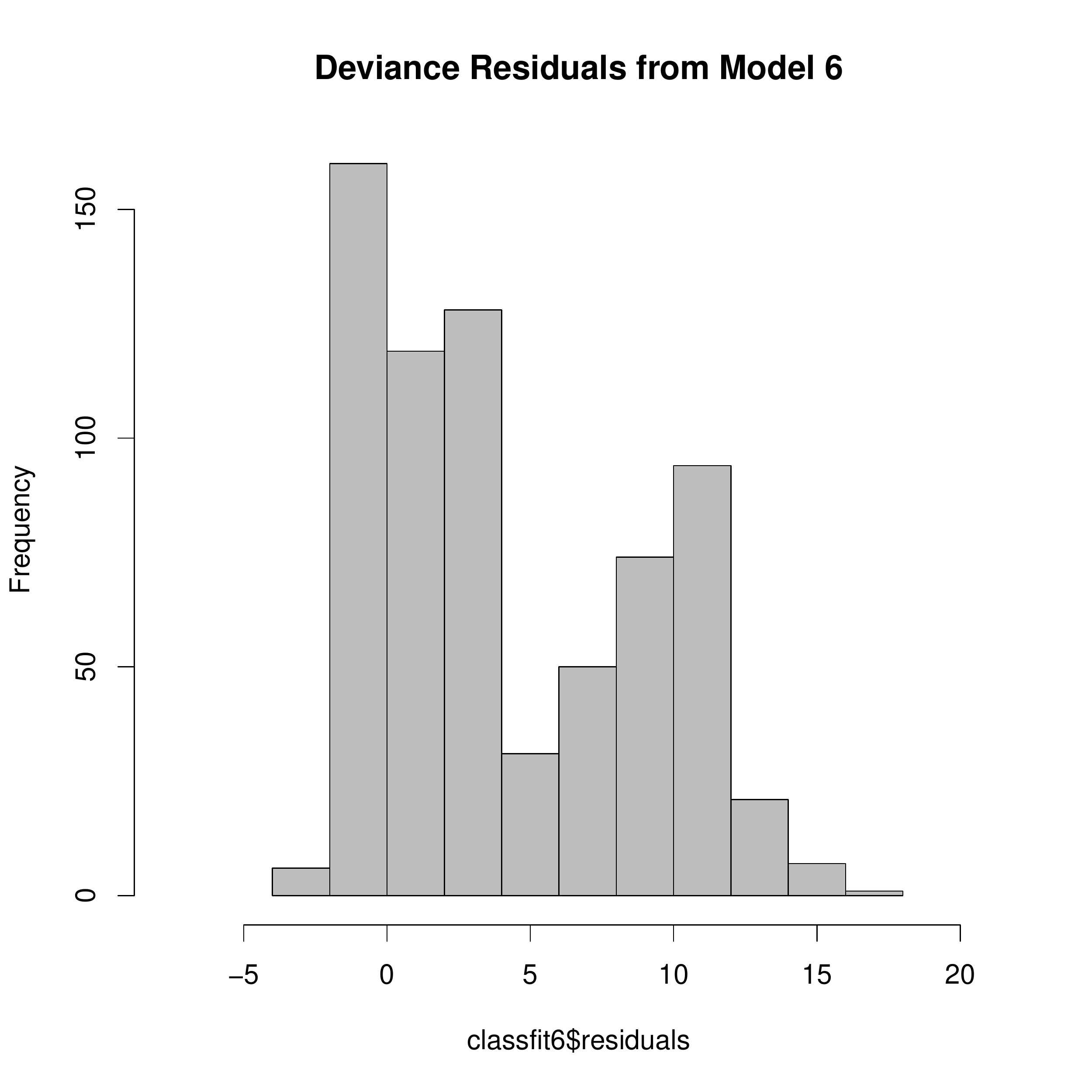}
\end{figure}

Despite not having a fixed null residual to evaluate against, we can still inspect the distribution of the deviance residuals.  Ideally, we would like them to be small and clustered near zero. Figure~\ref{fig.classhist} plots the histogram of the deviance residuals from model 6.  The distribution is clearly more ``lumpy" than that observed in Figure~\ref{fig.wtchist} for the corresponding the WTC model, suggesting that the classroom dyamics are less well-predicted on average than were the radio communications.  
\begin{lstlisting}
> #Plot the histogram of the deviance residuals from model 6
> hist(classfit6$residuals)

> #How well do we predict the exact event?
> mean(apply(classfit6$predicted.match,1,all))
[1] 0.3299566

> #How well do we predict either the sender or receiver of an event?
> mean(apply(classfit6$predicted.match,1,any))
[1] 0.5166425

> #How well do we predict each part of the event?
> colMeans(classfit6$predicted.match)
   FromId      ToId 
0.5050651 0.3415340 
\end{lstlisting}

\begin{figure}
\caption{Classification Accuracy of the Observed Ranks Under Model 6 with Prediction Thresholds Indicated at 0.05, 0.1, and 0.25 \label{fig.ecdf2}}
\centering
\includegraphics[width=\textwidth]{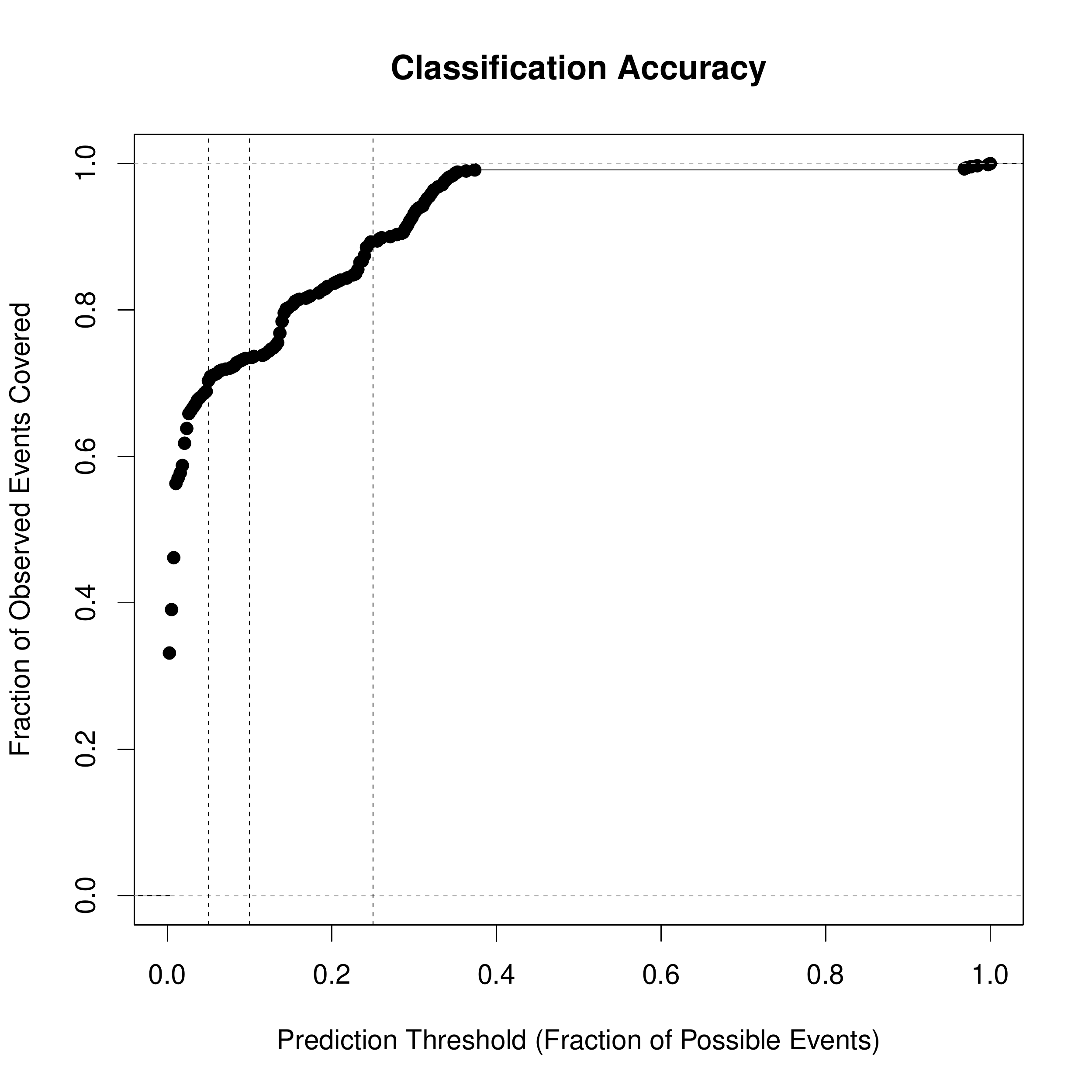}
\end{figure}

Evaluating how well the model predicts each event sheds additional light on these results.  On average, the model only predicts the event perfectly about 33\% of the time (still a remarkable performance, given the large number of possible events).  We do a bit better with getting at least one part of the event right, correctly classifying the sender or receiver about 50\% of the time (and we do much better at classifying senders than receivers over all, on average).  Moreover, inspection of the classification accuracy in Figure~\ref{fig.ecdf2} for this model shows substantial lag between the prediction threshold and fraction of the observed events covered by the model. By 25\% of the possible events transpiring, the model has only predicted 89\% of the observed events (compared with 98\% in the corresponding WTC case).

\begin{lstlisting}
> #Classification plot
> plot(ecdf(classfit6$observed.rank/(19*20)), xlab="Prediction Threshold (Fraction of Possible Events)",ylab="Fraction of Observed Events Covered",main="Classification Accuracy",xlim=c(0,1))
> abline(v=c(0.05,0.1,0.25),lty=2)

> #a comparative look at the 25th prediction threshold
> ecdf(classfit6$observed.rank/(19*20))(.25)
[1] 0.8929088

> ecdf(wtcfit6$observed.rank/(37*36))(.25)
[1] 0.983368
\end{lstlisting}

So, comparatively, it looks that our exact time relational event model of the classroom data isn't performing as well as our ordinal time relational event model of the WTC data.  We may be missing some important aspect of the relational event process in our model of the classroom conversation.  We can again examine the model ``surprise" superimposed on the time-aggregated network for clues about what may be going on. Here, because we lack a null residual, we'll define surprising events as those for which the observed event is not in the top 5\% of those predicted. 

\begin{lstlisting}
> #rank > 19 corresponds to the 5\% cut-off here
> surprising<-as.sociomatrix.eventlist(Class[classfit6$observed.rank>19,],20)

> #Plot the resulting surprising network
> gplot(surprising,edge.lwd=surprising^0.75,arrowhead.cex=log(as.edgelist.sna(surprising)[,3])+.25,vertex.col=ifelse(WTCPoliceIsICR,"black","gray"),vertex.cex=1.25,vertex.sides=ifelse(WTCPoliceIsICR,4,100),displayisolates=FALSE)
\end{lstlisting}

\begin{figure}
\caption{Time-Aggregated `Surprising' Events Network Under the Final Relational Event Model of McFarland's Classroom Data \label{fig.class2}}
\centering
\includegraphics[width=\textwidth]{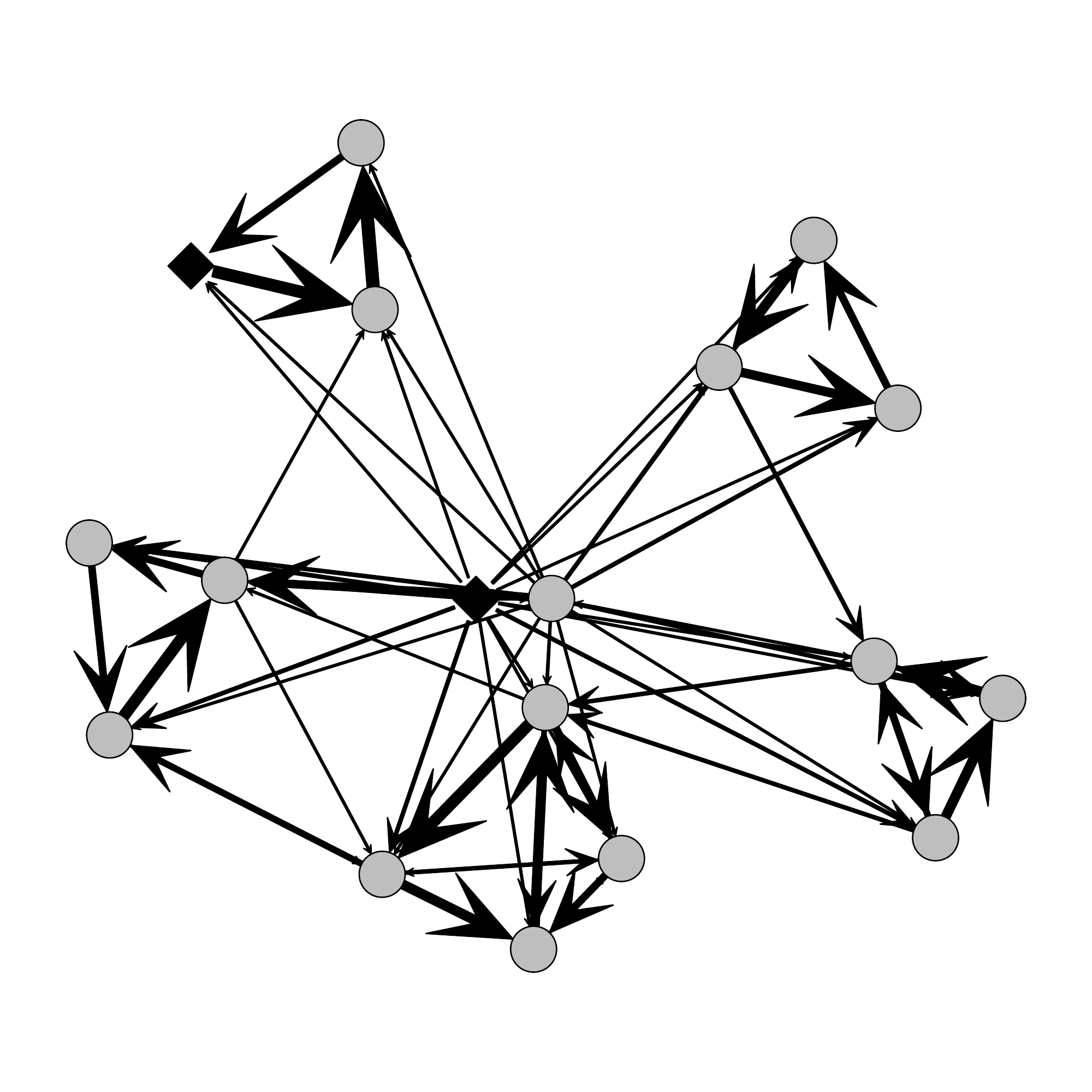}
{\flushleft \scriptsize Two isolates were suppressed for visualization.}
\end{figure}

The visualization in Figure~\ref{fig.class2} gives us more of a clue about what we're missing. Specifically, the presence of five distinct clusters represent the occurrence of various side discussions that are not well-captured by the current model.  This could be due to the fact that things like P-shift effects fail to capture simultaneous side-conversations (each of which may have its own set of turn-taking patterns), or to a lack of covariates that capture the enhanced propensity of subgroup members to address each other (such as students being in the same school club together).  Further elaboration could be helpful here.  On the other hand, we seem to be doing reasonably well at capturing the main line of discussion within the classroom, particularly vis-a-vis the instructors.  Whether or not this is adequate depends on the purpose to which the model is to be put; as always, adequacy must be considered in light of specific scientific goals.

\section{Conclusion}

A wide range of interaction processes---from radio communications to dominance contests---can be fruitfully studied within the relational event paradigm.  While arising as the short-duration limit of the dynamic network regime, the relational event regime has its own distinct properties and requires distinct treatment.  In particular, relational event dynamics are fundamentally about \emph{sequential relational structure,} rather than the \emph{simultaneous relational structure} that is the dominant concern within social network analysis.  In this and many other respects, theory and analysis of relational event dynamics owes as much to fields such as conversation analysis, event history analysis, and agent-based modeling as to conventional network analysis.  Relational event models are still fundamentally structural, however, and we stress that the approaches are complementary.  Indeed, where exact (or exactly ordered) data is available on relationship start and stop times, it is possible to model dynamic networks via a REM process whose events involve the creation and termination of edges.  Taking such a process to be fully latent---with only the state of the currently active edges observed at a small number of distinct points in time---leads one to a model family that is essentially similar to the framework of \citet{snijders:sm:2001}.  Likewise, temporally extensive relationships are often important covariates for relational event processes, allowing one to directly assess the impact of ongoing ties on social microdynamics.

Although we have focused here on some of the most basic types of REMs, more complex cases are also possible.  As noted, REMs for ``egocentric'' event data \citep{marcum&butts2015csse} can be powerful tools for modeling the responses of individuals to their local social environments, and are well-suited to the analysis of complex event series (with many event types) punctuated by exogenous events.  Hierarchical extensions to REMs \citep{dubois&2013hmre} allow for pooling of information across multiple event sequences while still allowing the dynamics of each sequence to differ from the others; this is particularly useful when studying many small groups, and/or when attempting to estimate covariate effects for attributes whose prevalence varies greatly from group to group.  Endowing REMs with latent structure also holds a host of opportunities, including the ability to infer latent \emph{interaction roles} directly from behavioral data \citep{dubois3&2013sbre}.  Given the breadth and flexibility of the approach, the prospects are good for many more developments in this area.

We close with the important reminder that no representation is fit for all purposes, nor is it intended to be.  Many relational analysis problems involve the modeling of ongoing relationships, and are better viewed through the lenses of static or dynamic network analysis.  Where one's focus is on micro-interaction or other processes involving discrete behaviors whose implications cascade forward through time, however, the relational event paradigm offers a powerful and statistically grounded alternative.



\singlespacing
\bibliography{rem}
\end{document}